\documentclass[12pt,a4paper]{article}
\usepackage[utf8]{inputenc}
\usepackage[T1]{fontenc}
\usepackage{jcappub}

\usepackage{xcolor}
\usepackage{amsmath}
\usepackage{amssymb}

\usepackage[textwidth=3cm,textsize=footnotesize]{todonotes}

\newcommand{\keV}{\,\text{keV}}
\newcommand{\MeV}{\,\text{MeV}}
\newcommand{\eV}{\,\text{eV}}
\newcommand{\GeV}{\,\text{GeV}}

\newcommand{\cL}{\mathcal{L}}
\newcommand{\lAngle}{\langle\!\langle}
\newcommand{\rAngle}{\rangle\!\rangle}

\newcommand{\disp}{\displaystyle}

\newcommand{\half}{\frac{1}{2}}

\newcommand{\eff}{{\rm eff}}
\newcommand{\osc}{{\rm osc}}
\newcommand{\res}{{\rm res}}
\newcommand{\OmegaDM}{\Omega_{\rm DM}}
\newcommand{\dec}{{\rm dec}}
\renewcommand{\l}{\left(}
\renewcommand{\r}{\right)}

\title{Scalar induced resonant sterile neutrino production in the
  early Universe}
\author[a]{F. Bezrukov,}
\author[b,c]{A. Chudaykin,}
\author[b,c]{D. Gorbunov}

\affiliation[a]{The University of Manchester, School of Physics and Astronomy,\\
  Oxford Road, Manchester M13 9PL, United Kingdom}
\affiliation[b]{Institute for Nuclear Research of the Russian Academy of Sciences,\\
  60th October Anniversary prospect 7a, Moscow 117312, Russia}
\affiliation[c]{Moscow Institute of Physics and Technology,\\
  Institutsky per. 9, Dolgoprudny 141700, Russia}

\emailAdd{Fedor.Bezrukov@manchester.ac.uk}
\emailAdd{chudy@ms2.inr.ac.ru}
\emailAdd{gorby@ms2.inr.ac.ru}

\date{\today}

\abstract{
  It has been recently suggested
  \cite{Bezrukov:2017ike,Bezrukov:2018wvd} that a cosmic scalar field
  can completely change the keV-scale sterile neutrino production in
  the early Universe.  Its effect may, for various parameter choices,
  either suppress sterile neutrino production and make moderate
  active-sterile mixing cosmologically acceptable, or increase the
  production and generate considerable dark matter component out of
  sterile neutrino with otherwise negligible mixing with SM.  In this
  paper we provide analytic estimates  complementing and providing details of the
  numerical calculations performed in \cite{Bezrukov:2018wvd} in the
  case of resonant amplification of the sterile neutrino
  production. We also discuss phenomenological and theoretical issues
  related to the successful implementation of this idea in fully
  realistic extensions of the Standard Model of particle physics.
}

\begin{document}

\begin{flushright}
	INR-TH-2019-021
\end{flushright}

\maketitle
\flushbottom

%%%%%%%%%%%%%%%%%%%%%%%%%%%%%%%%%%%%%%%%%%%%%%%%%%%%%%%%%%%%%%%%%%%%%%%%
\section{Introduction} 
\label{Sec:intro}

The numerous dark matter phenomena getting gradually accepted
\cite{Bertone:2016nfn} ask for either modification of General Relativity or extension of
the Standard Model of particle physics (SM). In the latter case of a
special interest are those extensions which are motivated by particle
physics itself, independently from cosmology. These extensions must
contain a new electrically neutral, stable at cosmological time-scales
particles and provide a mechanism operating in the early Universe to
produce them in a sufficient amount. That is to yield a relative
contribution of the new species (several in case of multicomponent dark
matter) $\Omega_M\simeq 0.27$ to the present energy density of
the Universe.

One of the well-motivated candidates to form one of the dark matter
components is sterile neutrinos --- singlet with respect to SM gauge
group fermions, inducing neutrino masses via mixing with the SM (or
active) neutrinos. In this way neutrino oscillations
 get explained, see e.g.\ \cite{Gorbunov:2014efa,Bilenky:2016pep}. The
sterile-active mixing makes sterile neutrino unstable. Consequently, the
sterile neutrinos decay via weak interactions into light SM particles,
with the decay into three active neutrinos being always
open. However, since the weak decay rate goes with the decaying particle
mass $M_X$ as $\Gamma_W\propto M_X^5$, sufficiently light sterile
neutrinos may be long-lived enough to survive till the present, with
the lifetime exceeding the Universe age $\tau_U\simeq 1.4\times
10^{10}$\,y. This is one of the necessary conditions of a 
 dark matter component. It selects keV-mass scale as cosmologically
viable, provided the sterile-active mixing is enough for the sterile
neutrino contributing to active neutrino mass, for
details see e.g.\,\cite{Adhikari:2016bei,Abazajian:2017tcc}. However,
at present, this chain of arguments is misleading actually, as we
explain below.

Remarkably, the instability suggests a
specific signature of the sterile neutrino dark matter:
it radiatively decays into active neutrino and photon. This two
body decay implies a peak feature in the Galactic photon spectrum 
at the energy of half of the sterile neutrino mass ($X$-rays for
the keV mass-scale). Extensive searches for this peak (which relative width is
of order $10^{-3}$ due to the Doppler shift of an object bound by
the Galaxy gravity force) revealed no signal so far, and severely
constrained the sterile neutrino model parameter space. These searches
place upper limits on the active-sterile mixing at a given sterile
neutrino mass, under the assumption of all the dark matter being formed by the
sterile neutrino only. These bounds do not depend on a mechanism of
the sterile neutrino production operating in the early Universe, but
if sterile neutrino is only a subdominant dark matter component, the
limits are correspondingly weaker.

The relic sterile neutrinos must be produced in the early Universe at
some stage of its expansion in the amount right enough to form the
entire dark matter component, or at least its fraction. The
production mechanisms, based on active-sterile neutrino oscillations
in the primordial plasma provide with sterile neutrino of momenta
proportional to the plasma temperature. Hence, the keV-scale neutrinos
form so-called {\it warm dark matter}\,\cite{Adhikari:2016bei}, a dark matter
variant subjecting to specific constraints from the cosmic structure
formation \cite{Schneider:2016uqi} and from phase space
density \cite{Boyarsky:2008ju,Gorbunov:2008ka}. Together with X-ray
searches, the most recent one is \cite{Roach:2019ctw}, 
these constraints exclude the simplest mechanism of
sterile neutrino production via the oscillations in the primordial plasma
 \cite{Dodelson:1993je} and leave only small regions in the model
parameter space cosmologically viable in the situation when the
primordial plasma is asymmetric with respect to the lepton charge 
\cite{Shi:1998km}. While the latter case is still allowed (e.g.\ within
the $\nu$MSM setup), {\it the sterile
neutrino there does not contribute enough to active neutrino masses, } and
one may ask for other mechanisms, either extensions or alternatives to
those mentioned above. {\it To our knowledge none of the suggested
in literature mechanisms simultaneously explains the active neutrino
masses via mixing and produces the full dark matter component consistent with
  cosmological and astrophysical bounds discussed above.}    

The sterile neutrinos of keV-mass range are most efficiently produced
in the oscillations at temperature of about 100\,MeV
\cite{Dodelson:1993je}. Therefore, to amplify or suppress the production via
oscillations one can respectively increase or decrease the effective
sterile-active mixing in the primordial plasma at this epoch. As we
pointed out in Refs.\,\cite{Bezrukov:2017ike,Bezrukov:2018wvd}, the
latter effect can be achieved by coupling the sterile neutrino to a
scalar field with specific homogeneous evolution in the expanding
Universe \footnote{The same framework has been recently invoked to suppress the cosmological production of eV-scale sterile neutrinos, see Ref. \cite{Farzan:2019yvo,Cline:2019seo}.}. With a constant scalar field this coupling can make the
sterile neutrino either massless or superheavy in the interesting
epoch, thus naturally preventing the sterile neutrino
production\,\cite{Bezrukov:2017ike}. Alternatively, rapidly
oscillating scalar can resonantly amplify the neutrino
oscillations\,\cite{Bezrukov:2018wvd} and enhance the sterile neutrino
production in comparison with the standard case \cite{Dodelson:1993je}. The periodically varying scalar field can induce a parametric resonance phenomenon which dramatically increases the amplitude of neutrino oscillations regardless of active-sterile mixing. This
mechanism has been studied numerically in
Ref.\,\cite{Bezrukov:2018wvd}, which showed that the small (as
compared to the standard case) sterile-active mixing may be sufficient
to produce enough dark matter. This observation justifies a further
search for the peak-like signature in the $X$-rays with the
next-generation telescopes such as eRosita and ART-XC on the base of
Spektr-RG platform \cite{Pavlinsky:2008ecy,Merloni:2012uf}.

Parametric resonance phenomenon in systems coupled to
the oscillating background is not a novel in particle physics. In one of 
the pioneering studies \cite{Pusch:1982ps} devoted to neutron–antineutron oscillations 
in a periodically varying magnetic field a strong enhancement of the probability amplitude was explored. Later, amplification of spin-flavor neutrino oscillations in electromagnetic fields of various configurations was found in Ref. \cite{Egorov:1999ah}. In Ref. \cite{Dvornikov:2000mt} 
the method for the investigation of the neutrino evolution equation solution
near the resonance point was suggested. A novel approach to study neutrino oscillations in the presence of general rapidly varying fields without assumptions about the strength of the time-varying external field was developed in Ref. \cite{Dvornikov:2004ms}. Instead of solving the evolution equation directly the author has derived the new effective Hamiltonian which described the evolution of the averaged neutrino wave function. In contrast to Ref. \cite{Dvornikov:2000mt}, the elaborated method was beyond the perturbation treatment and applicable for arbitrary strength of the external field. However, a new approach requires a large frequency of the time-varying background compared to that of the system at the absence of any external force. For this reason, it is instructive to explore the resonance dynamics of neutrino oscillations in the presence of oscillating background in different assumption about external conditions. In this sense, our analysis is complementary to one in Ref. \cite{Dvornikov:2004ms} and can be used to examine the parametric resonance phenomenon at a different choice of model parameters.

In this paper we develop the analytic approach to describe the
resonant dynamics, estimate the amount of and calculate the spectrum
of the produced dark matter sterile neutrinos. This study represents a logical extension of the previous analysis \cite{Bezrukov:2018wvd}. We corroborate the numerical treatment of Ref.\,\cite{Bezrukov:2018wvd} with rigorous analytical approach which allows to extend the scope of our mechanism and provides proper analytic understanding of the dependence of cosmological predictions on the model parameters. In particular, we generalise the framework of Ref. \cite{Bezrukov:2018wvd} in the presence of thermal environment which allows to examine resonance dynamics in the high temperature regime.

%It fill the gaps in the numerical analysis of Ref.\,\cite{Bezrukov:2018wvd}, and allow for proper analytic understanding of the dependence of cosmological predictions on the model parameters. It generalises our approach in the presence of thermal environment which allows to examine resonance dynamics in the high temperature regime.

The paper is organized as follows. In
Section\,\ref{Sec:resonance} we describe the oscillating sterile-active neutrino
system in the expanding Universe as the Schroedinger equation with
external field and find the resonance solution. In
Section\,\ref{Sec:spectrum} we suggest an analytic approximation to the
spectrum of resonantly produced sterile
neutrinos. Section\,\ref{Sec:dark_matter} contains the analytic
estimate of the total amount of produced in this way sterile neutrino
dark matter. We apply here all the cosmological and astrophysical
constraints on the model parameters. Section\,\ref{Sec:issues} 
is devoted to discussion of various issues to be resolved on the way
of implementation of the developed mechanism of sterile neutrino
production in the realistic extensions of the SM (e.g.\ seesaw type I
with Majorana field or secluded scalar sector). We summarize our
results in Conclusion, Section\,\ref{Sec:sum}.
Some lengthy analytic estimates are presented in Appendix \ref{Sec:second_order} for
convenience. 

%%%%%%%%%%%%%%%%%%%%%%%%%%%%%%%%%%%%%%%%%%%%%%%%%%%%%%%%%%%%%%%%%%%%%%%%%%%%%%

\section{Neutrino resonance induced by oscillating background}
\label{Sec:resonance}
%consists of
%We consider SM active neutrino sterile neutrinos evolution 
Our model Lagrangian contains three essential parts: 
\begin{equation}\label{eq:total}
\cL=\cL_N+\cL_{\phi N}+\cL_\phi\,.
\end{equation}
The first one describes oscillations between active neutrino $\nu$ and
its right-handed sterile counterpart $N$ (fermion singlet with respect
to the SM gauge group) in vacuum. It reads 
\begin{equation}\label{eq:L}
\cL_N = i \bar \nu \hat\partial \nu+i \bar N \hat\partial N + \frac{M}{2} \bar{N^c} N
+ m_D \bar \nu_a N  + \text{h.c.}.
\end{equation}
where $m_D$ is the Dirac mass appeared after the electroweak transition and equals 
\begin{equation}\label{dirac_mass}
m_D=\theta M
\end{equation}
where $\theta$ is the mixing angle in vacuum (i.e. at present, in the
cosmological context) and $M$ is the bare mass of sterile neutrino.
The Dirac term implies the second non-zero mass eigenvalue of the
$\nu$-$N$ system, the active neutrino mass of order $m_a\sim
m_D^2/M\sim\theta^2 M$. 
The second ingredient of our model \eqref{eq:total}
is the real scalar field $\phi$ (the SM singlet as
well) which feebly couples to sterile neutrino via Yukawa like interaction
\begin{equation}\label{eq:LNh}
\cL_{\phi N} = \frac{f}{2}\, \phi \bar{N^c} N + \text{h.c.}
\end{equation}
The Yukawa term \eqref{eq:LNh} gives rise to the time-dependent contribution 
\begin{equation}
\label{time-dependent-mass_only}
M_N=f\phi(t) 
\end{equation}
to the effective sterile neutrino mass 
\begin{equation}
\label{Meff}
M_\eff\equiv M+M_N.
\end{equation}
It implies that the scalar field treated as an external force here may
control the neutrino oscillations in the early Universe. The corresponding
impact is determined by dynamic of the scalar sector. To simplify the
analysis we will use the theory of free massive scalar field
\begin{equation}\label{eq:phi}
\cL_{\phi} = \half(\partial_\mu\phi)^2 + \half m_\phi^2\phi^2
\end{equation}
In this framework, at early times the scalar field is frozen and the sterile neutrinos can be very heavy \eqref{time-dependent-mass_only}
\begin{equation}\label{M-initial}
  M_{N,i}=f\phi_i.
\end{equation}
When the Universe expansion rate (i.e. the Hubble parameter) drops
below $H_{\osc}\simeq m_\phi$ the scalar field begins to oscillate
that results in the variable sterile neutrino mass
contribution\,\eqref{time-dependent-mass_only} oscillating with frequency $m_\phi$ \eqref{time-dependent-mass_only}
\begin{equation}
  \label{neutrino-mass-fall}
  M_N=M_A\sin m_\phi t.
\end{equation}
and decreasing amplitude
\begin{equation}\label{MA}
M_A\equiv M\left(
    \frac{h T^3}{h_{e}^{\vphantom{3}} T_e^3}
  \right)^{\!1/2}\,.
\end{equation}
Hereafter we assume the Universe at radiation domination and
parameterize the amplitude\,\eqref{MA} so, that at the temperature
$T=T_e$ it coincides in value with the bare mass $M_N$. 

In what follows we study the dynamics of active-to-sterile neutrino
transitions in the presence of oscillating scalar field coupled to
the sterile massive state. Sum of \eqref{eq:L} and \eqref{eq:LNh} can be
rewritten in a more contracted form \eqref{time-dependent-mass_only},
\eqref{Meff}
\begin{align}\label{flav_bas}
\begin{split}
\cL_N+\cL_{\phi N}=i\bar{\mathcal{N}}_L \hat\partial \mathcal{N}_L&+\half\bar{\mathcal{N}}_L^c
\begin{pmatrix}
0 & m_D \\ m_D & M_\eff
\end{pmatrix}
\mathcal{N}_L+\text{h.c.},\\
\mathcal{N}_L&\equiv
\begin{pmatrix}
\nu\\
N^c
\end{pmatrix}
\end{split}
\end{align}
or in terms of Majorana fields $\mathcal{N}\equiv\l\mathcal{N}_L+\mathcal{N}_L^c\r/\sqrt{2}$ it reduces to
\begin{align}\label{flav_bas1}
\begin{split}
\cL_N+\cL_{\phi N}=i\bar{\mathcal{N}} \hat\partial \mathcal{N}&+\bar{\mathcal{N}}
\begin{pmatrix}
0 & m_D \\ m_D & M_\eff
\end{pmatrix}
\mathcal{N}.
\end{split}
\end{align}

Massive part of~\eqref{flav_bas1} can be diagonalized by the following orthogonal transformation
\begin{align}\label{3:transf_matr}
\begin{split}
\mathcal{O}=&
\begin{pmatrix}
\cos\theta_\eff & \sin\theta_\eff\\ -\sin\theta_\eff & \cos\theta_\eff
\end{pmatrix}
\end{split}
\end{align}
where effective time-dependent mixing $\theta_\eff$ is defined through
\begin{equation}\label{3:transf_matr_thet}
\tan\theta_\eff=\frac{2m_D}{M_\eff+\sqrt{M_\eff^2+4m^2_D}}.
\end{equation}

Upon transformation $\mathcal{N}=\mathcal{O}\mathcal{V}$ \eqref{3:transf_matr} Lagrangian \eqref{flav_bas1} arrives at
\begin{align}\label{mass_bas}
\begin{split}
\cL_N+\cL_{\phi N}=i\bar{\mathcal{V}}\hat\partial\mathcal{V}+&i\mathcal{V}^\dag
\begin{pmatrix}
0 & \dot\theta_\eff \\ -\dot\theta_\eff & 0
\end{pmatrix}
\mathcal{V}+\bar{\mathcal{V}}
\begin{pmatrix}
m_1 & 0 \\ 0 & m_2
\end{pmatrix}
\mathcal{V},\\
m_{1,2}&=\frac{M_\eff}{2}\l1\mp\sqrt{1+4\frac{m_D^2}{M_\eff^2}}\r,
\end{split}
\end{align}
and squared mass difference of \eqref{mass_bas} reads
\begin{equation}\label{3:transf_matr_m}
\Delta m^2=M_\eff^2\sqrt{1+\frac{4m_D^2}{M_\eff^2}}
\end{equation}

To describe evolution of the two-level system \eqref{mass_bas} in the
presence of oscillating background \eqref{Meff}, \eqref{neutrino-mass-fall} 
we employ matrix form of the
Schrödinger equation. Evolution of massive neutrino states in
vacuum \footnote{Implementation of plasma effects is rather straightforward,
  see for details Sec. \ref{subsec:plasma}.} under the most general
assumptions is governed by \eqref{mass_bas}
%Substituting a diagonal contribution
\begin{equation}\label{Schrod_mass_bas}
i\frac{\partial}{\partial t}
\begin{pmatrix}
\nu_1 \\ \nu_2
\end{pmatrix}
=\begin{pmatrix}
-\Delta_{0,\eff}/2 & -i\dot\theta_\eff \\ i\dot\theta_\eff & \Delta_{0,\eff}/2
\end{pmatrix}
\begin{pmatrix}
\nu_1 \\ \nu_2
\end{pmatrix}
\end{equation}
where $\nu_1$, $\nu_2$ denote wave functions of massive states and we also defined the effective rate of neutrino oscillations in vacuum \eqref{3:transf_matr_m}
\begin{equation}\label{3:osc_rate0}
\Delta_{0,\eff}\equiv\frac{\Delta m^2}{2p}
\end{equation}

Analysis of neutrino oscillations via \eqref{Schrod_mass_bas} is rather involved. Framework \eqref{Schrod_mass_bas} can be simplified significantly under several assumptions. First, if effective angle \eqref{3:transf_matr_thet} changes slowly enough, the massive states $\nu$ can be treated as stationary and possible convertions $\nu_1\leftrightarrow\nu_2$ can be neglected. The corresponding adiabacity condition reads 
\begin{equation}\label{3:adiab_constr0}
\frac{|\dot\theta_\eff|}{\Delta_{0,\eff}}\ll1\,. 
\end{equation}
Assuming \eqref{3:adiab_constr0} the original framework \eqref{Schrod_mass_bas} for massive states can be rewritten in the flavour basis using transformation $(\nu_1,\nu_2)^{\rm T}=\mathcal{O}^{-1}(\psi_1,\psi_2)^{\rm T}$ \eqref{3:transf_matr} as follows
\begin{equation}\label{3:Schrod_mass_bas2}
i\frac{\partial}{\partial t}
\begin{pmatrix}
\psi_1 \\ \psi_2
\end{pmatrix}
=\mathcal{H}
\begin{pmatrix}
\psi_1 \\ \psi_2
\end{pmatrix}
\end{equation}
where $\psi_1$, $\psi_2$ refer to the flavour basis and the effective
Hamiltonian reads
\begin{equation}\label{3:qHamilton}
\mathcal{H}=\frac{\Delta_{0,\eff}}{2}\left(\begin{array}{cc}
-\cos2\theta_\eff & \sin2\theta_\eff\\
\sin2\theta_\eff & \cos2\theta_\eff
\end{array}
\right).
%,\;\text{где}\quad \Delta_0 = \frac{\Delta m^2}{2 p}
\end{equation}

Second, in addition to \eqref{3:adiab_constr0} one can require smallness of effective mixing \eqref{3:transf_matr_thet}. Assuming
\begin{equation}\label{angle_constr}
\frac{2m_D}{M_\eff}\ll1\,,
\end{equation}
we found simplified forms of oscillating parameters \eqref{3:transf_matr_m}, \eqref{3:transf_matr_thet}, \eqref{dirac_mass}
\begin{align}\label{3:angle_of_time}
\begin{split}
\Delta m^2&\approx M_\eff^2\,,\\
\sin^2\theta_\eff&\approx\frac{m_D}{M_\eff}\,.
%=\frac{\theta z}{z+\sin m_\phi t}
\end{split}
\end{align}

Assuming \eqref{3:angle_of_time} and exploiting the following notations 
\begin{equation}\label{3:osc_rate}
\Delta_{0,\eff}\equiv2\beta(z+\sin m_\phi t)^2,\qquad \beta\equiv\frac{M_A^2}{4p}\,,\qquad z\equiv\frac{M}{M_A}\,,
\end{equation}
eq. \eqref{3:Schrod_mass_bas2} reduces to
\begin{equation}\label{3:Schrod_mass_bas3}
i\frac{\partial}{\partial t}
\begin{pmatrix}
\psi_1 \\ \psi_2
\end{pmatrix}
=\beta(z+\sin m_\phi t)^2\begin{pmatrix}
-1 & \frac{\disp 2\theta z}{\disp z+\sin m_\phi t} \\  \frac{\disp 2\theta z}{\disp z+\sin m_\phi t} & 1
\end{pmatrix}
\begin{pmatrix}
\psi_1 \\ \psi_2
\end{pmatrix}.
\end{equation}

Additional assumptions \eqref{3:adiab_constr0}, \eqref{angle_constr},
whose validity is exhaustively discussed in Appendix
\ref{Sec:second_order}, allow one to reduce the rather complicated
framework \eqref{Schrod_mass_bas} to a more straightforward dynamics
governed by \eqref{3:Schrod_mass_bas3}. To solve the latter we
introduce the basis flavour states in the absence of mixing: 
$|\psi_a\rangle=\begin{pmatrix} \psi_1^{(0)} \\ 0 \end{pmatrix}$ and
$|\psi_s\rangle=\begin{pmatrix} 0
\\ \psi_2^{(0)} \end{pmatrix}$. These states can be easily founded as
solutions of the diagonal part of the Schrodinger
equation~\eqref{3:Schrod_mass_bas3}, namely $\psi_{1,2}^{(0)}=e^{\pm
  i\beta\int_0^t(z+\sin m_\phi\zeta)^2d\zeta}$.

%We introduce the state vector $|\psi(t)\rangle\equiv\begin{pmatrix} \psi_1 \\ \psi_2 \end{pmatrix}$ which obeys~\eqref{3:Schrod_mass_bas3}. $|\psi_a\rangle=\begin{pmatrix} \psi_1^{(0)} \\ 0 \end{pmatrix}$ and $|\psi_s\rangle=\begin{pmatrix} 0 \\ \psi_2^{(0)} \end{pmatrix}$ are the basis flavour states in the absence of mixing. These states can be easily founded as solutions of the diagonal part of Schrödinger equation~\eqref{3:Schrod_mass_bas3}, namely  $\psi_{1,2}^{(0)}=e^{\pm i\beta\int_0^t(z+\sin m_\phi\zeta)^2d\zeta}$. 

So, the state vector $|\psi(t)\rangle\equiv\begin{pmatrix} \psi_1 \\ \psi_2 \end{pmatrix}$ at any moment can be presented in the following form
\begin{equation}\label{stat_vec}
|\psi(t)\rangle=y_1(t)|\psi_a\rangle+y_2(t)|\psi_s\rangle,
\end{equation}
where $y_1(t)$ and $y_2(t)$ are corresponding flavour amplitudes
\begin{align}\label{ampl}
\begin{split}
y_1(t)=\langle\psi_a|\psi(t)\rangle,\\
y_2(t)=\langle\psi_s|\psi(t)\rangle.
\end{split}
\end{align}
We are interested in the so-called appearance probability of the
sterile state, so we put $|\psi(0)\rangle=|\psi_a\rangle$. In this
case $y_1(t)$ and $y_2(t)$ describe survival and transition
amplitudes, respectively.

Substituting \eqref{stat_vec} into \eqref{3:Schrod_mass_bas3} gives
rise to the following system of equations on flavour amplitudes
\eqref{ampl}
\begin{equation}\label{ampl_eq_syst}
\begin{cases}
\frac{\disp\partial y_1(t)}{\disp\partial t}=-2i\beta z\theta(z+\sin m_\phi t)e^{-2i\beta\int_0^t(z+\sin m_\phi\zeta)^2d\zeta}y_2(t)\\
\frac{\disp\partial y_2(t)}{\disp\partial t}=-2i\beta z\theta(z+\sin m_\phi t)e^{2i\beta\int_0^t(z+\sin m_\phi\zeta)^2d\zeta}y_1(t)
\end{cases}
\end{equation}
with initial conditions $y_1(0)=1$, $y_2(0)=0$. 

In what follows we work in the limit of small energy transfer between
the active and sterile states, that means  
\begin{equation}\label{y_small}
|y_2(t)|\ll 1\,. 
\end{equation}
In this approximation the number density of active neutrinos reduces insignificantly during all oscillation period $y_1(t)\simeq 1$. This assumption allows one to obtain the solution of \eqref{ampl_eq_syst} in the following form
\begin{equation}\label{sol_int}
y_2(t_1)\simeq-2i\beta z\theta\int_0^{t_1}(z+\sin m_\phi t)e^{2i\beta\int_0^t(z+\sin m_\phi\zeta)^2d\zeta}dt\,.
\end{equation}

The outer integration in \eqref{sol_int}, over $t$, can be carried out
with a help of stationary phase method. This method is valid only for
significantly large ratios $2\beta/m_\phi$. We assume that to apply
this strategy one needs at least the equality 
\begin{equation}\label{3:beta_m}
m_\phi\leqslant\beta
\end{equation}
to be true. Further calculations can be simplified in the regime of
large scalar field amplitude, that requires $M\ll M_A$ or, given 
\eqref{3:osc_rate}, 
\begin{equation}\label{3:z_small}
z\ll1\,.
\end{equation}

\subsection{Resonant solution}
\label{subsec:phase_method}

Denoting  
\begin{align}\label{denot}
\begin{split}
h(t)&\equiv2\int_0^t(z+\sin m_\phi\zeta)^2d\zeta,\\
g(t)&\equiv z+\sin m_\phi t\,,
\end{split}
\end{align}
the condition $h'(t_l)=0$ brings the following stationary points $t_l$
\begin{equation}
(z+\sin m_\phi t_l)=0
\end{equation}
or,  adopting \eqref{3:z_small}, 
\begin{equation}\label{stat_points}
m_\phi t_l\approx-(-1)^{l}z+\pi l\,.
\end{equation}

We start with the estimate of the contribution of the second order stationary point $t_l\in[0,t_1]$ to \eqref{sol_int}. To achieve that one should accomplish the following integration \eqref{denot}
\begin{align}\label{sol_prel}
\begin{split}
\int_{t_l-\delta}^{t_l+\delta} g(t)e^{i\beta h(t)}dt\approx
g'(t_l)e^{i\beta h(t_l)}\int_{t_l-\delta}^{t_l+\delta}
&(t-t_l)e^{i\frac{1}{6}\beta(t-t_l)^3h'''(t_l)}dt\,. 
%&g'(t_l)\frac{1}{3}\Gamma\l\frac{2}{3}\r\l\frac{6}{\beta|h'''(t_l)|}\r^{2/3}e^{i\beta h(t_l)+i\frac{\pi}{3}{\rm sgn}h'''(t_l)}
\end{split}
\end{align}
For that we make use of the asymptotic methods that yields the following reference integrals \cite{Erdelyi}
\begin{align}\label{sol_erdai}
\begin{split}
\int_0^\delta t\,e^{ixt^3}dt&\sim\frac{1}{3}\,\Gamma\l\frac{2}{3}\r x^{-2/3}e^{i \pi/3},\qquad\,\, x\rightarrow+\infty\\
\int_{-\delta}^0 t\,e^{ixt^3}dt&\sim-\frac{1}{3}\,\Gamma\l\frac{2}{3}\r x^{-2/3}e^{-i \pi/3},\quad x\rightarrow+\infty
\end{split}
\end{align}
Using \eqref{sol_erdai}, the asymptotic expansion of \eqref{sol_prel} in the limit $2\beta/m_\phi\rightarrow+\infty$ reduces to 
\begin{equation}\label{sol_point0}
\int_{t_l-\delta}^{t_l+\delta} g(t)e^{i\beta h(t)}dt\sim
ig'(t_l)\,\frac{\sqrt{3}}{3}\,\Gamma\l\frac{2}{3}\r\l\frac{6}{\beta|h'''(t_l)|}\r^{2/3}e^{i\beta
  h(t_l)}. 
\end{equation}
Assuming the amplitude \eqref{sol_point0} and approximations 
\begin{align}\label{denot1}
\begin{split}
h'''(t_l)&=4m_\phi^2\l\cos 2m_\phi t_l-z\sin m_\phi t_l\r\approx4m_\phi^2\\
g'(t_l)&=m_\phi\cos m_\phi t_l\approx (-1)^l\,m_\phi
\end{split}
\end{align}
we write down the contribution of $l$-th stationary point to the integral \eqref{sol_int}
\begin{align}\label{sol_point}
\left.y_{2}\right|_{l}\!=\!2z\theta(-1)^l\sqrt{3}\,\Gamma\!\l\frac{2}{3}\r\!\l\!\frac{\beta}{12m_\phi}\!\r^{1/3}\!\!\exp\left\{i\frac{2\beta}{m_\phi}\l\frac{\disp\pi
  l(1\!+\!2z^2)}{\disp 2}\!+\!2z(1\!-\!(-1)^l)\r\!\right\}. 
\end{align}

To estimate the intagral \eqref{sol_int} over large periods of time
one needs to sum up contributions \eqref{sol_point} over the all
stationary points \eqref{stat_points} encountered in the time interval
$[0,t_1]$, namely $\sum\left.y_{2}\right|_{l}$. To attain this goal,
we split $\sum\left.y_{2}\right|_{l}$ into odd and even parts as
follows, 
\begin{align}
%\begin{split}
\sum&\left.y_{2}\right|_{l}=\sum_{l=2k}\left.y_{2}\right|_{l}+\sum_{l=2k+1}\left.y_{2}\right|_{l},\nonumber\\
%\label{3:sum1}
\label{sum}
\sum_{l=2k}\left.y_{2}\right|_{l}=2z\theta\sqrt{3}\,\Gamma\!\l\frac{2}{3}\r&\l\!\frac{\beta}{12m_\phi}\!\r^{1/3}\sum_{k=0}^{l_{\rm max}/2}\!\exp\left\{\disp i\frac{2\beta}{m_\phi}\cdot\pi k(1\!+\!2z^2)\right\},\\
%\label{3:sum2}
\sum_{l=2k+1}\!\!\left.y_{2}\right|_{l}=-2z\theta\sqrt{3}\,\Gamma\!\l\frac{2}{3}\r&\l\!\frac{\beta}{12m_\phi}\!\r^{1/3}\sum_{k=0}^{l_{\rm max}/2}\!\exp\left\{i\frac{2\beta}{m_\phi}\l\frac{\pi (2k\!+\!1)}{2}(1\!+\!2z^2)\!+\!4z\r\!\right\},\nonumber
%\end{split}
\end{align}
where $l_{\rm max}$ refers to the final time $t_1$,
see~\eqref{stat_points}, as 
\begin{equation}\label{lmax}
l_{\rm max}\approx\frac{m_\phi t_1}{\pi}\,. 
\end{equation} 

The sums in \eqref{sum} are geometric progressions with the common
ratio $\exp\{i\frac{2\beta}{m_\phi}\pi(1+2z^2)\}$. The growing mode in
the solution $\left|\sum\left.y_{2}\right|_{l}\right|$ refers to the
following condition:  $\\\beta(1+2z^2)/m_\phi\in \mathbb{N}$ or
\eqref{3:osc_rate}, that is 
%$\exp\{i\frac{2\beta}{m_\phi}\pi(1+2z^2)\}=1$
\begin{equation}\label{k_def}
%n\equiv\frac{\beta(1+2z^2)}{m_\phi}\in \mathbb{N}
\frac{M_A^2+2M^2}{4p_n}=nm_\phi, \quad n\in \mathbb{N}\,.
\end{equation}
The condition \eqref{k_def} guarantees a parametric resonance in the
two-level system when the mass of one state is modulated by
oscillating background. The integer $n$ parameterizes diversity of the
parametric resonances in such a system. This situation refers to the most prominent conversions $\nu_a\rightarrow\nu_s$.
% mostly enhanced oscillations
%the most pronounced growth of $|y_2(t)|$ (in fact infinite growth, see below) and the most prominent conversions $\nu_a\rightarrow\nu_s$.

Eventually, assuming \eqref{lmax}, \eqref{k_def}, \eqref{3:osc_rate}
we simply reduce \eqref{sum} to the leading order in $z$
\eqref{3:z_small} as  
\begin{equation}\label{sol_answer}
|y_{\rm2,lin}(t)|\approx\left|\sum\left.y_{2}\right|_{l}\right|\approx0.65\,\theta\, m_\phi t\frac{M}{M_A}n^{1/3}
\begin{cases}
|\sin4\frac{M}{M_A}n|, \quad \text{for even}\,\, n\\
|\cos4\frac{M}{M_A}n|, \quad \text{for odd}\,\, n
\end{cases}
\end{equation}

\begin{figure}[!t]
\centering
  \includegraphics[width=0.46\linewidth]{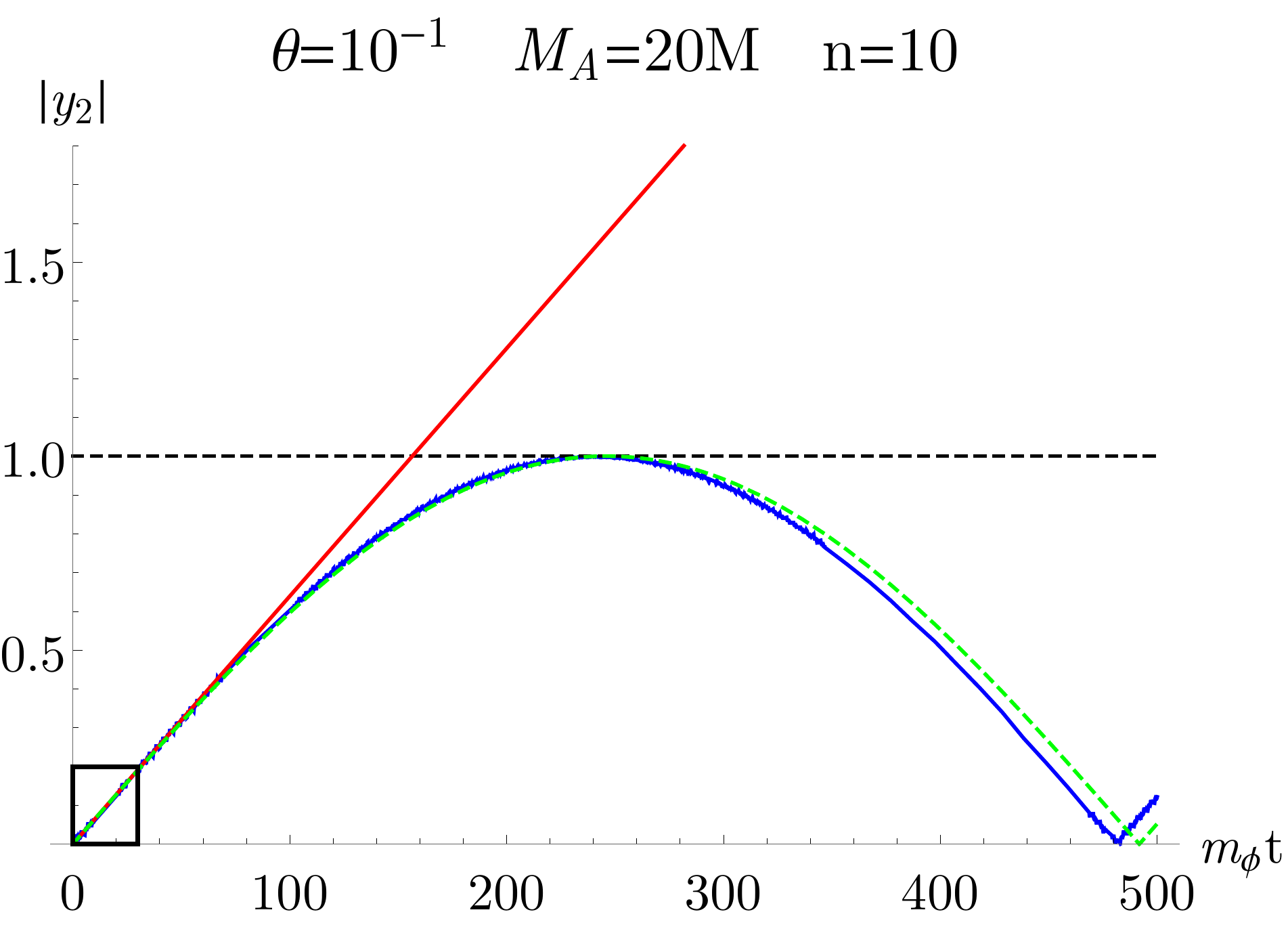}
  \includegraphics[width=0.46\linewidth]{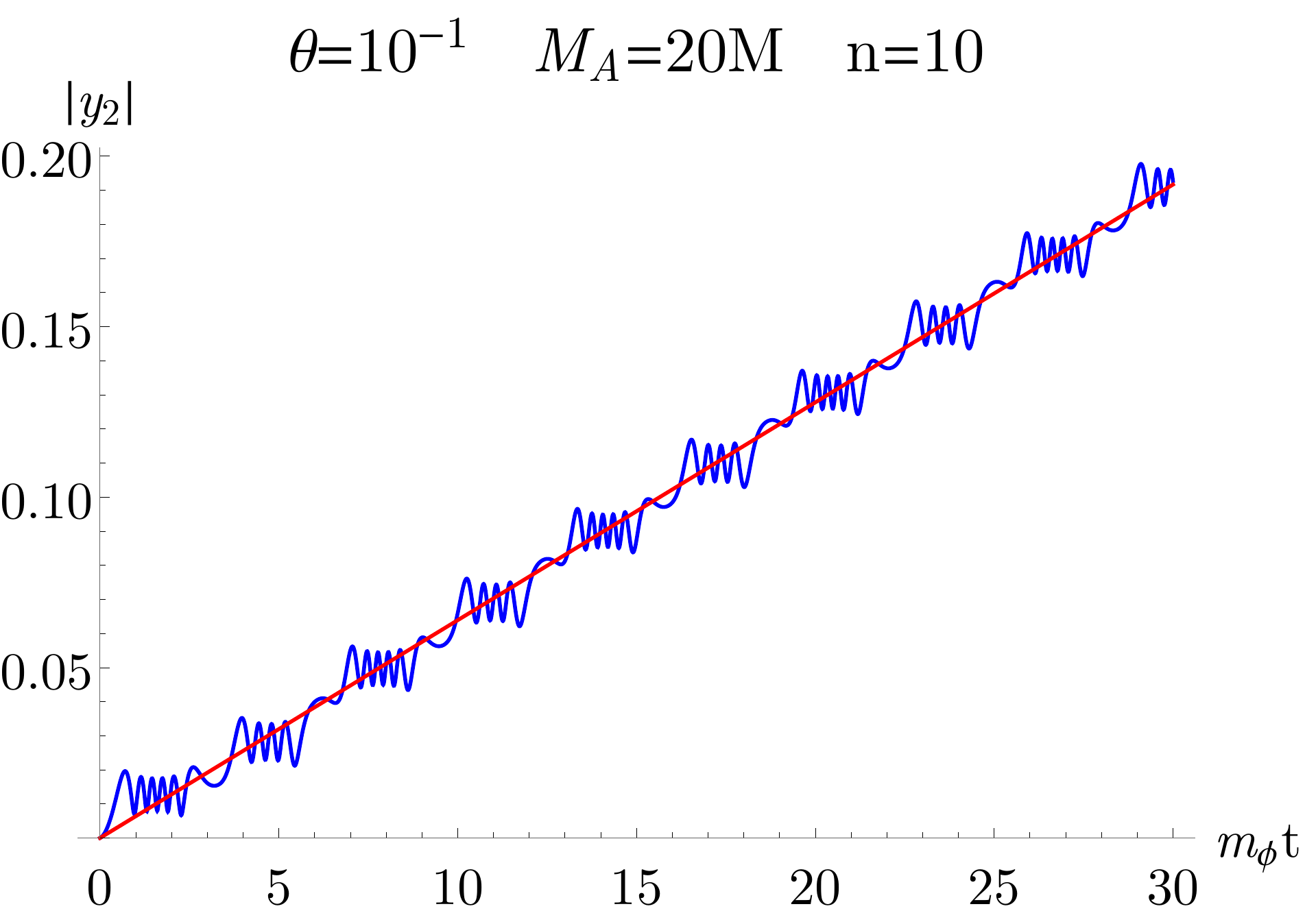}
  	\includegraphics[width=0.46\linewidth]{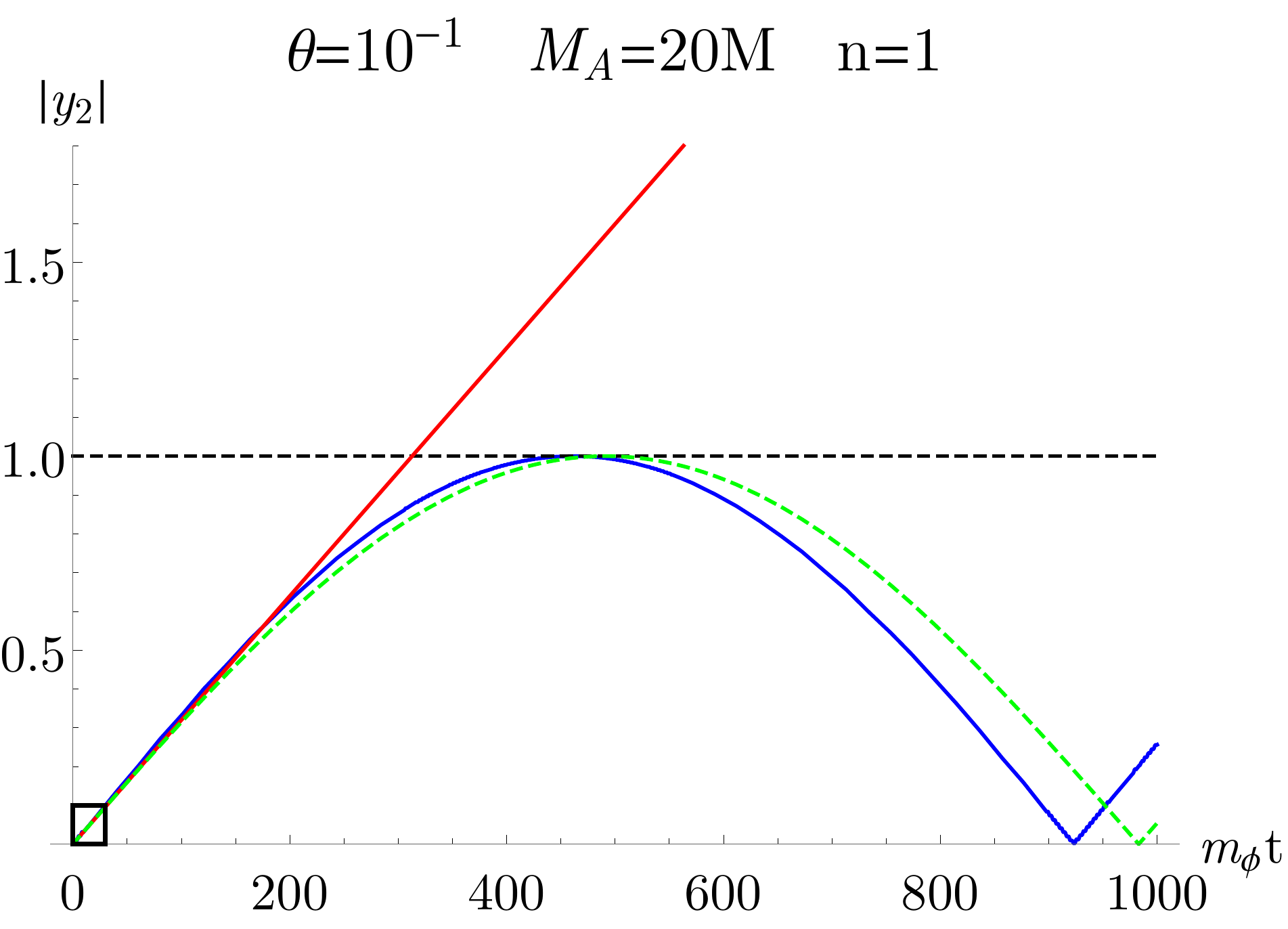}
  	\includegraphics[width=0.46\linewidth]{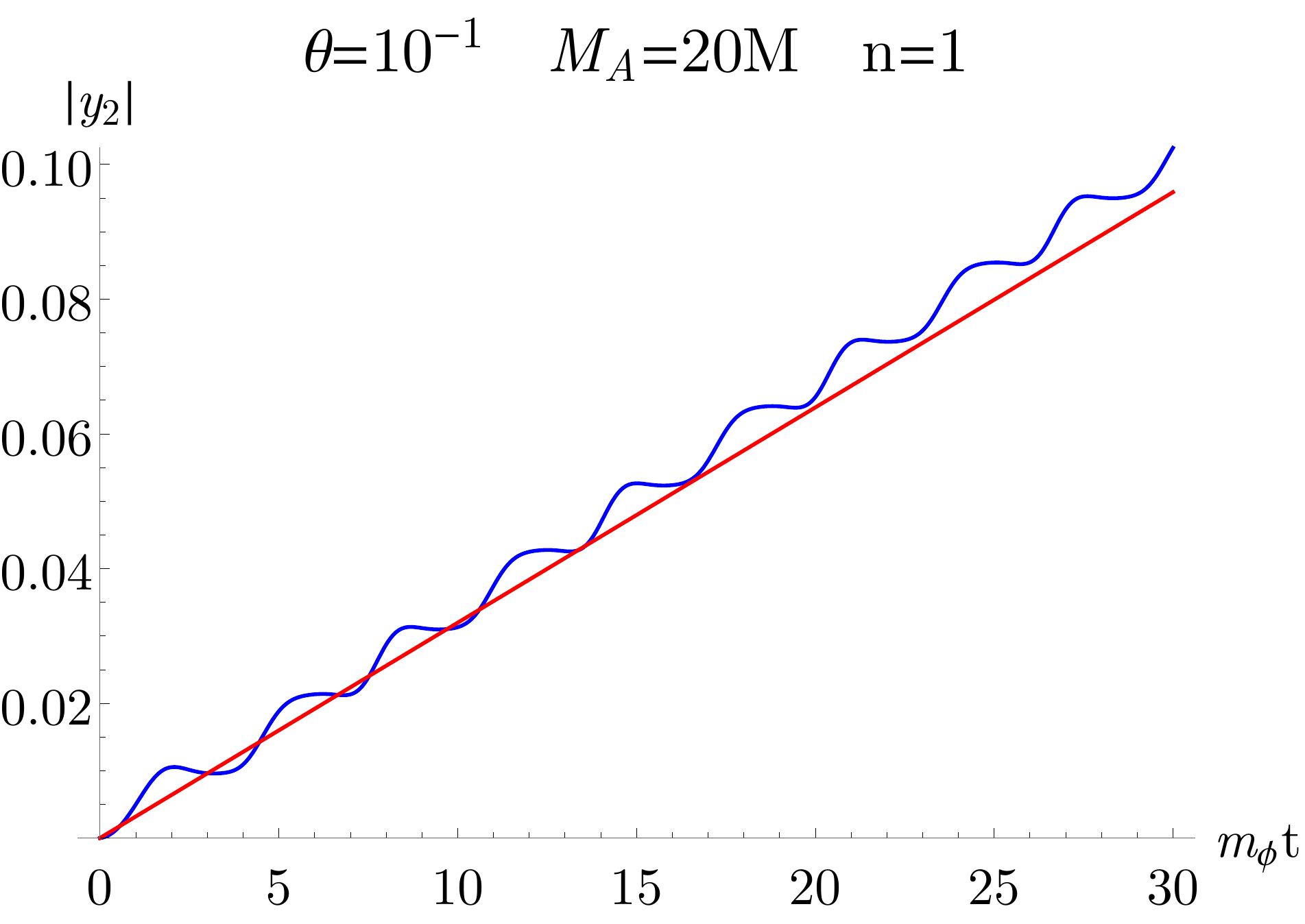}
  \caption{The blue line represents the numerical solution of
    equations \eqref{ampl_eq_syst}. The red line depicts our
    analytical approximation \eqref{sol_answer} in the regime
    \eqref{y_small}. The green dashed line refers to our phenomenological result \eqref{3:sol_answer1}.}
  \label{fig:3}
\end{figure}

In order to demonstrate the applicability of our theoretical approach,
we confront the analytical approximation \eqref{sol_answer} with numerical
solution of equations \eqref{ampl_eq_syst} for resonances with
various values of $n$, see \eqref{k_def}. Fig. \ref{fig:3} shows that our
estimate (red line) in the regime \eqref{y_small} reproduces the numerical
solution (blue line) very precisely for $n=10$ and more modestly in
case $n=1$. However, this nice accordance is broken if one examines
the resonant behaviour at $y_2(t)\lesssim 1$. Given this reason, one needs
to find a proper solution of \eqref{ampl_eq_syst} which does not rely on
\eqref{y_small}. To achieve this goal we examine, for particular sets
of variables, one phenomenological
dependency $|\sin y_{\rm 2,lin}(t)|$ (green dashed line) against
numerical solution (blue line) in Fig. \ref{fig:3}. The performed analysis
reveals that the first oscillation peak of the numerical solution is
approximated by this phenomenological model reasonably well. Thus, a
proper solution in the case of resonance \eqref{k_def} which does not rely on
\eqref{y_small} is given by \eqref{sol_answer}
\begin{equation}\label{3:sol_answer1}
y_{\rm 2,res}(t)\approx\left|\sin\l\frac{\omega_\res\cdot t}{2}\r\right|,
\end{equation}
where 
\begin{equation}\label{omega_res}
\omega_\res\approx1.3\,\theta\, m_\phi\frac{M}{M_A}n^{1/3}
\begin{cases}
|\sin4\frac{M}{M_A}n|, \quad \text{for even}\,\, n\\
|\cos4\frac{M}{M_A}n|, \quad \text{for odd}\,\, n
\end{cases}
\end{equation} 
%Finally, appropriate solution $y_{\rm 2,res}$ in resonance \eqref{k_def} reads \eqref{sol_answer}

%confront the numerical solution of equations \eqref{ampl_eq_syst} with our analytical approximation \eqref{sol_answer} valid at \eqref{y_small}. and result of one phenomenological model for different choice of parameters \eqref{k_def} in Fig.~\ref{fig:3}.

%However, in region $|y_{\rm 2,lin}(t)|\lesssim 1$ our estimate \eqref{sol_answer} becomes unreliable. In Fig.~\ref{fig:3} one phenomenological result $\left|\sin y_{\rm2, lin}(t)\right|$ is also depicted. We reveal herein that the first oscillation peak of the numerical solution \eqref{ampl_eq_syst} is approximated very well by this model. It allows to determine a frequency of proper solution by

%We emphasize that \eqref{sol_answer} gives an inreliable result when approximation of small energy transfer \eqref{y_small} does not hold. 

%As far as $|y_{\rm 2,lin}(t)|\ll 1$, analytic approximation \eqref{sol_answer} reproduces numerical solution \eqref{ampl_eq_syst} very precisely. However, in region $|y_{\rm 2,lin}(t)|\lesssim 1$ our estimate \eqref{sol_answer} becomes unreliable. In Fig.~\ref{fig:3} one phenomenological result $\left|\sin y_{\rm2, lin}(t)\right|$ is also depicted. We reveal herein that the first oscillation peak of the numerical solution \eqref{ampl_eq_syst} is approximated very well by this model. It allows to determine a frequency of proper solution by
%in a more precise manner by
%provides appropriate result during long evolution times. It allows us to  Assuming  
%\begin{equation}
%y_{\rm2,lin}\l\frac{\pi}{\omega_{\rm res}}\r\approx\frac{\pi}{2}
%\end{equation}
%or~\eqref{sol_answer}

%In what follows we need $y_{\rm 2,res}$ at $n=1$. Accuracy of asymptotic methods applied in this case can be traced in Fig. \ref{fig:33}.

Our phenomenological result \eqref{3:sol_answer1} manifests a good
resemblance with that of the numerical approach for any integer $n$, see
\eqref{k_def}. It justifies the application of stationary phase method and
reveals its efficiency to describe the parametric resonance in the two-level
system \eqref{ampl_eq_syst}.

\subsection{Width of resonance}
\label{subsec:width}

%about evolution of the system undergoing is the width of corresponding resonance.
The other important characteristics of a parametric resonance is its
width.  To obtain a range of momenta for which the flavour amplitudes
deviate from their resonant values moderately, we calculate
\eqref{sol_int} in some vicinity of $n$-resonance \eqref{k_def}
provided by
\begin{equation}\label{alpha_def}
\frac{M_A^2+2M^2}{4p_n}=(n+\alpha)m_\phi\,,
%\frac{\beta(1+2z^2)}{m_\phi}=n+\alpha,\quad n\in \mathbb{N},\quad\alpha<1
%,\quad\alpha\ll1
\end{equation}
where $\alpha\ll 1$. 
Using the stationary phase method elaborated in Sec. \ref{subsec:phase_method} we reduce \eqref{sum} to the following amplitude \eqref{lmax}, \eqref{alpha_def}
\begin{equation}\label{sol_answer_alpha}
|y_{2,\alpha}(t)|\approx0.65\,\theta\,m_\phi\frac{M}{M_A}n^{1/3}\frac{2\pi\left|\sin\l\frac{m_\phi\alpha}{2}t+\pi\alpha\r\right|}{|\sin\pi\alpha|}
\begin{cases}
|\sin(4\frac{M}{M_A}n+\frac{\pi\alpha}{2})|, \quad \text{for even}\,\, n\\
|\cos(4\frac{M}{M_A}n+\frac{\pi\alpha}{2})|, \quad \text{for odd}\,\, n
\end{cases}
\end{equation}
As expected, in the extremely small vicinity of resonance,
i.e. at $\alpha\rightarrow0$, our result \eqref{sol_answer_alpha} reproduces the linear resonant solution \eqref{sol_answer} under assumption of \eqref{y_small}. 

To find the relevant value of parameter $\alpha$ which governs
\eqref{alpha_def}, we introduce the absolute value of amplitude
$y_2(t)$
\begin{equation}\label{y_max}
{\rm max}\,|y_{2}|\approx\left|y_{2,\alpha}\l\frac{\pi}{m_\phi\alpha}\r\right|.
\end{equation}
Assuming $\alpha\ll1$ and $2\theta\frac{M}{M_A}n^{1/3}\ll {\rm max}\,|y_{2}|$ we can express $\alpha$ through ${\rm max}\,|y_{2}|$ in the following form \eqref{sol_answer_alpha}
\begin{equation}\label{alpha}
\alpha\approx\frac{1}{{\rm max}\,|y_{2}|}\,\frac{\omega_\res}{m_\phi}\,.
%\alpha\approx2z\theta_0\l\frac{n}{12}\r^{1/3}\sqrt{3}\,\Gamma\l\frac{2}{3}\r\times\l\frac{2}{\pi}\r\frac{1}{|y_{\rm2, max}|}
%\begin{cases}
%|\sin4zn| \quad \text{for even}\,\, n\\
%|\cos4zn| \quad \text{for odd}\,\, n
%\end{cases}
\end{equation}

Then, adopting \eqref{y_max}, \eqref{alpha} in the vicinity of
resonance \eqref{alpha_def}, the solution \eqref{sol_answer_alpha} can be presented as
\begin{equation}\label{3:sol_answer_alpha}
|y_{2,\alpha}(t)|\approx{\rm
  max}\,|y_{2}|\times\left|\sin\l\frac{\omega\cdot t}{2}\r\right|, 
\end{equation}
where
\begin{equation}\label{omega}
\omega\approx\frac{\omega_\res}{{\rm max}\,|y_{2}|}\,.
%\omega=2zm_\phi\theta_0\l\frac{n}{12}\r^{1/3}\sqrt{3}\,\Gamma\l\frac{2}{3}\r\times\l\frac{2}{\pi}\r\frac{1}{|y_{\rm2, max}|}
%\begin{cases}
%|\sin4zn| \quad \text{for even}\,\, n\\
%|\cos4zn| \quad \text{for odd}\,\, n
%\end{cases}
\end{equation}

\begin{figure}[!htb]
	\centering
	\includegraphics[width=0.47\linewidth]{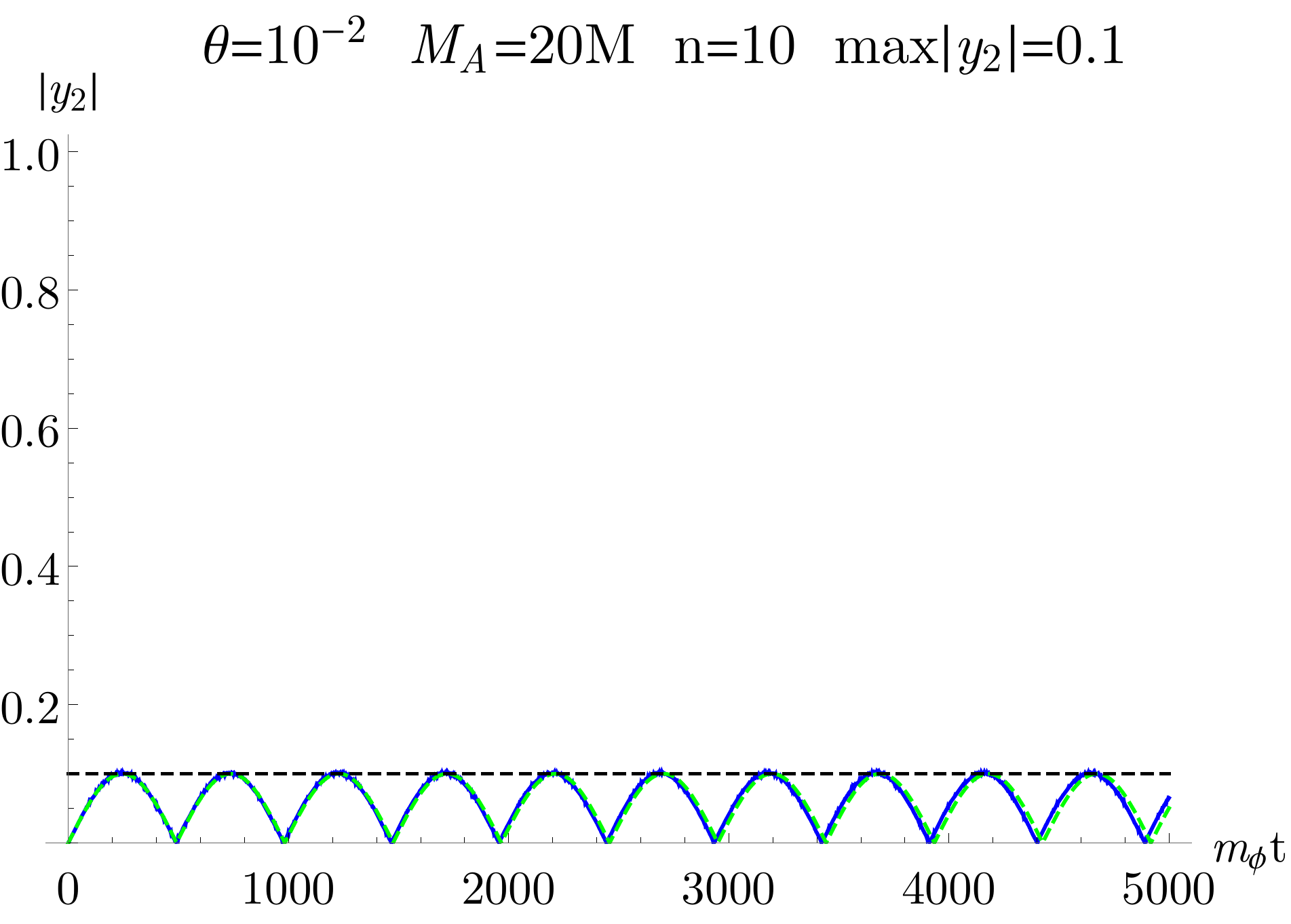}
	\includegraphics[width=0.47\linewidth]{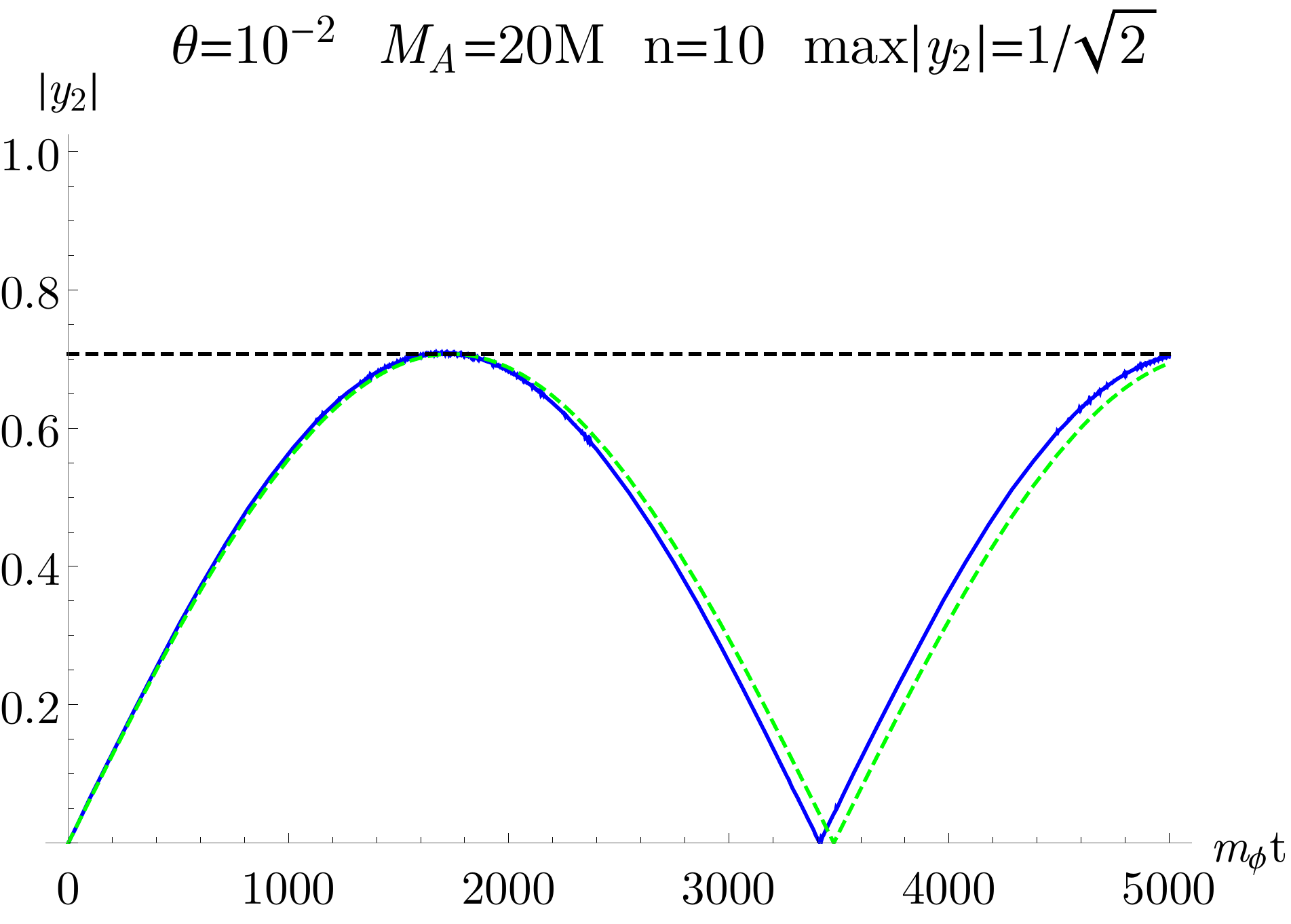}
	\includegraphics[width=0.47\linewidth]{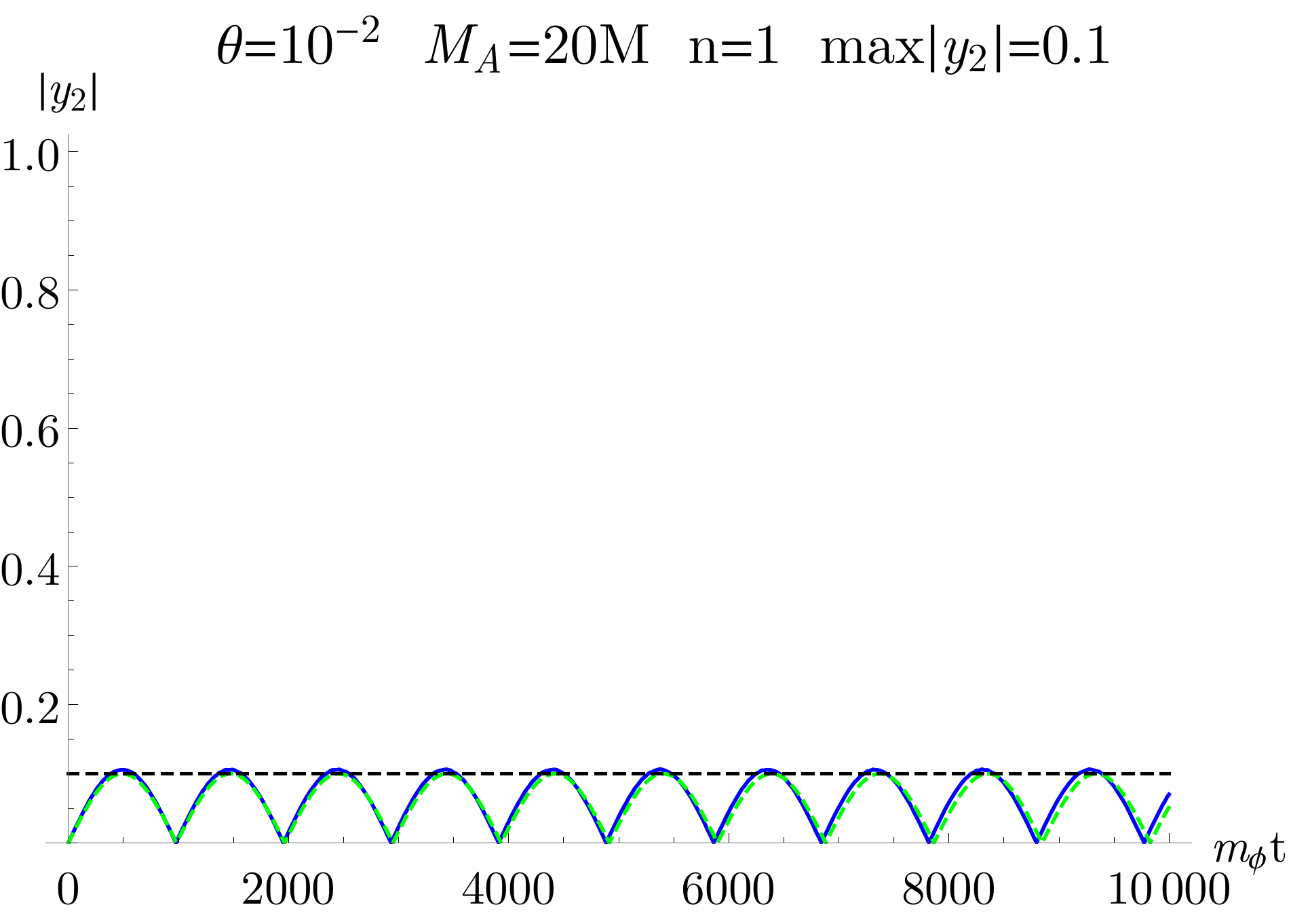}
	\includegraphics[width=0.47\linewidth]{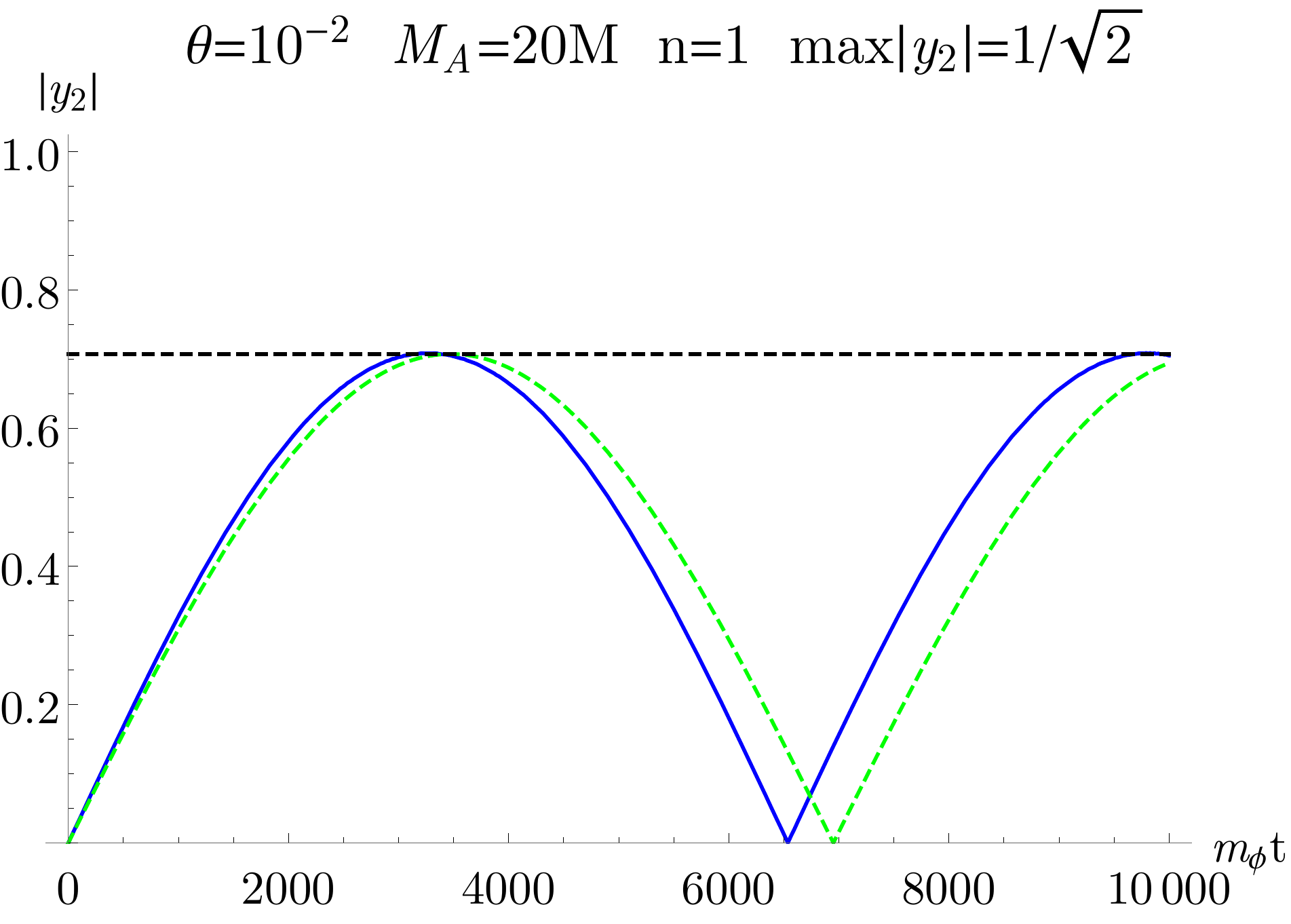}
	\caption{The blue line represents the numerical solution of
          equations~\eqref{ampl_eq_syst} on the amplitude ${\rm
            max}\,|y_{2}|$. The green dashed line depicts our
          estimate~\eqref{3:sol_answer_alpha} of the same amplitude
          ${\rm max}\,|y_{2}|$. The black dashed line shows the chosen value ${\rm max}\,|y_{2}|$ in each case.}
	\label{fig:4}
\end{figure}

%approximations \eqref{y_max} and \eqref{alpha} holds only in
%It proves the empirical dependency chosen in \eqref{3:sol_answer1}.
We recall that our results obtained above rely on the small energy
transfer condition \eqref{y_small}. However, the final outcome
\eqref{3:sol_answer_alpha} in the limit ${\rm
  max}\,|y_{2}|\rightarrow1$ matches the evolution at the resonance
\eqref{3:sol_answer1}. It suggests that our result
\eqref{3:sol_answer_alpha} can be applied for arbitrary value of
$y_{2}(t)$ if only its amplitude ${\rm max}\,|y_{2}|$ is treated as a
free parameter and its relation with \eqref{alpha} is not exploited.
To verify this statement, we confront our estimate \eqref{3:sol_answer_alpha}
with the numerical solution \eqref{ampl_eq_syst} of the same amplitude
for various values of ${\rm max}\,|y_{2}|$ and $n$. Fig.~\ref{fig:4}
reveals that our approach (dashed green line), indeed, can be applied
even for a big enough ${\rm max}\,|y_{2}|$.
%\eqref{omega} describes the frequency of exact solution near the resonance~\eqref{alpha_def} very precisely.

%The frequency of~\eqref{sol_answer_alpha} on large time scales equals $\omega=m_\phi\alpha$ or exploiting \eqref{alpha}, \eqref{omega_res}
%Interesting to note that frequency \eqref{omega} of solution \eqref{sol_answer_alpha} in extremely small vicinity of resonance ${\rm max}\,|y_{2}|\rightarrow1$ precisely reproduce that of resonance solution \eqref{omega_res}.

Eventually, we define a range of momenta, $\Delta p_n\equiv |p-p_n|$,
for which ${\rm max}\,|y_{2}|>1/\sqrt{2}$ holds via \eqref{alpha_def}: 
\begin{equation}\label{yk_width}
\Delta p_n= \frac{\alpha|_{1/\sqrt{2}}}{n}p_n\,,
\end{equation}
where $\alpha|_{1/\sqrt{2}}$ relates to \eqref{alpha} at ${\rm
  max}\,|y_{2}|=1/\sqrt{2}$ \footnote{We stress that relation between
  $\alpha$ and ${\rm max}\,|y_{2}|$ \eqref{alpha} for such a big ${\rm
    max}\,|y_{2}|$ becomes inaccurate. The reason of that consists in
  assumption \eqref{y_small} which we adopted to derive the solution in
  vicinity of resonance \eqref{sol_answer_alpha}. However, the
  corresponding discrepancy for ${\rm max}\,|y_{2}|=1/\sqrt{2}$
  appears not critical since this only leads to extra pre-factor in
  \eqref{alpha} which will be properly accounted for in the subsequent fitting procedure of the spectrum in Sec. \ref{Sec:spectrum}.}. Thus, the width of resonance reads \eqref{yk_width}
\begin{equation}\label{width}
\frac{\Delta p_n}{p_n}\simeq\frac{\sqrt{2}}{n}\frac{\omega_\res}{m_\phi}\,.
%\simeq\frac{4}{n}\frac{M}{M_A}\theta
\end{equation}

For our choice of parameters \eqref{3:z_small} the resonance
\eqref{k_def} is {\it narrow}, $\Delta p_n/p_n<1$. It implies
that the neutrino conversion in the resonance might be inefficient once
one considers expansion of the Universe. The latter effect gives rise
to a particular feature in the spectrum of sterile neutrinos produced in
resonance that is the subject of Sec. \ref{Sec:spectrum}.  But before
moving forward we study the resonance behaviour in the presence of a dense
environment composed of the thermally-populated SM particles.
%that is the subject of the next session.  

%\begin{equation}\label{alpha_half}
%\alpha|_{1/\sqrt{2}}=\sqrt{2}\frac{\omega_{\rm res}}{m_\phi}
%\end{equation}

\subsection{Plasma effect}
\label{subsec:plasma}

The results obtained above are based on the vacuum oscillation
framework. However, at high temperatures the active neutrino propagates
through dense cosmological plasma that affects oscillations. This
effect may be crucial for the resonant behaviour at high
temperatures. Given this reason, we generalise our results obtained in
Sec. \ref{subsec:phase_method} and \ref{subsec:width} in the presence
of matter.

The forward scattering of active neutrino in the thermal bath induces
the following effective Hamiltonian \eqref{3:qHamilton}
\begin{equation}\label{3:qHamilton_plasma}
\mathcal{H}_\eff=\mathcal{H}-\left(\begin{array}{cc}
V & 0\\
0 & 0
\end{array}
\right), 
\end{equation}
where $V$ denots effective potential for active neutrinos in the
plasma. We consider the mixing of sterile state only with electron
neutrino and vanishing initial primordial lepton number, so
\cite{Abazajian:2001nj}
\begin{equation}\label{2:VeffG}
V\approx63\times\Gamma\,, \qquad  \Gamma\approx1.27\times G_F^2T^4 p
\,. 
\end{equation}
%In what follows we examine how the plasma influence on a coherent evolution of neutrino states.
%reveal that matter effect can be subdominant in some special cases. 

In what follows we consider significantly high scalar field values and
examine matter effects in the following reasonable assumption,  
\begin{equation}\label{V_small}
V\ll\frac{M_A^2}{4p}\,.
\end{equation}
Then one can apply the stationary phase method elaborated in
Sec. \ref{subsec:phase_method} directly to effective Hamiltonian
\eqref{3:qHamilton_plasma} and obtain the following resonant condition
in the plasma, 
\begin{equation}\label{k_def_plasma}
\frac{M_A^2+2M^2}{4p_n}+V=nm_\phi\,, \quad n\in\mathbb{N}\,.
\end{equation}
This relation generalizes the resonance condition in vacuum
\eqref{k_def} and provides a bit different neutrino momentum at the
resonance $p_n$ assuming \eqref{V_small}. In turn, the resonance
solution in the presence of matter  matches that in the vacuum
\eqref{3:sol_answer1} and exhibits the same frequency
\eqref{omega_res}. This means that in assumption of \eqref{V_small}
the plasma effects do not affect the resonant behavior and introduce
only a slight offset in the resonance condition
\eqref{k_def_plasma}. The contribution of this displacement is
not significant either, if one enquires about the average resonant
momentum $p_n$ provided by \eqref{k_def_plasma} in the assumption of \eqref{V_small}.
%which is the matter of next session.  
%\footnote{Leading order correction to \eqref{omega_res} in the presence of plasma refers to $\sqrt{V4p}/M_A\ll1$ and can be neglected here, see \eqref{V_small}.}.

%Interpretation of result \eqref{k_def_plasma} is straightforward. In case $V<\frac{\disp M^2}{\disp 2p_n}$ plasma brings only sub-dominant contribution and evolution resembles the vacuum regime, see \eqref{k_def}. When $V>\frac{\disp M^2}{\disp 2p_n}$ plasma effect becomes relevant and leads to a small offset of resonance in accordance with \eqref{k_def_plasma}. 
%We stress that $\frac{\disp M^2}{\disp 2p}\ll\frac{\disp M_A^2}{\disp 2p}$, see \eqref{3:z_small}. 
%In fact, approach developed here and 

%We emphasize that the most delicate assumptions of our approach are \eqref{3:z_small} and \eqref{V_small}. 

Finally, using the numerical analysis based on \eqref{3:Schrod_mass_bas2}, \eqref{3:qHamilton_plasma}, \eqref{3:qHamilton} with precise
oscillation parameters \eqref{3:transf_matr_thet}, \eqref{3:transf_matr_m} we verified that our analytical approach elaborated in Sec. \ref{subsec:phase_method} and \ref{subsec:width} works reasonably well until 
\begin{align}\label{cond1}
&5M\lesssim M_A\,,\\
\label{cond2}
&5V\lesssim \frac{M_A^2}{4p}
\end{align} 
become valid, see \eqref{MA}, \eqref{2:VeffG}.

%%%%%%%%%%%%%%%%%%%%%%%%%%%%%%%%%%%%%%%%%%%%%%%%%%%%%%%%%%%%%%%%%%%%%%%%%%%%%%%%%

\section{Spectrum with narrow resonance}
\label{Sec:spectrum}
% To examine this possibility we introduce conformal momentum $y_n$ related to $n$-resonance \eqref{k_def} at temperature $T$

Expansion of the Universe eventually moves the system through the
region , where the resonance condition for a particular mode
\eqref{k_def} is fulfilled, that can thereby provide with meaningful
sterile neutrino abundance. In particular, redshift effect gives rise
to time-dependency of resonant conformal momentum $y_n$ passing
through $n$-resonance at temperature $T$ \eqref{k_def}
\begin{equation}\label{yk}
y_n\equiv\frac{p_n}{T}\approx\frac{M_A^2}{4Tm_\phi}\frac{1}{n}\,.
\end{equation}
The whole band of the width $\Delta y_n=\Delta p_n/T$ \eqref{width} moves through  a given resonant momentum $y_n$ \eqref{yk} over time 
\begin{equation}\label{res_dur}
\delta t_n=\frac{\Delta
  y_n}{\dot{y}_n}=\frac{1}{n}\frac{\omega_\res}{\sqrt{2}m_\phi H}\,. 
\end{equation}
If this period is shorter than typical time of resonant oscillations $\omega_\res^{-1}$, the resonance becomes ineffective ("narrow"). This happens at high temperature when 
\begin{equation}\label{res_slow}
\frac{\sqrt{2}m_\phi H}{\omega_\res^2}n > 1\,.
\end{equation}

% on the spectrum at high momenta (temperatures)
The l.h.s. of \eqref{res_slow} describes the efficiency of
$n$-resonance at temperature $T$ in the expanding Universe. Hence,
this quantity can be used to parametrize the spectrum of sterile
neutrinos produced in the case of "narrow" resonance. First, we
recover the asymptotic behaviour of sterile neutrino distribution function
in several limits of \eqref{res_slow} on theoretical grounds. Second,
we resort to numerical analysis and build the interpolating solution for
sterile neutrino spectrum valid for any reasonable value of parameter
\eqref{res_slow}.

In the static limit, $n\sqrt{2}m_\phi H/\omega_\res^2\ll 1$, the
amount of sterile neutrino reaches the Fermi--Dirac distribution after
typic time $\omega_\res^{-1}$ \eqref{3:sol_answer1},
so \footnote{Hereinafter, we assume that the plasma effect is significant
  on the typical time-scale of resonant oscillations $\sim
  1/\omega_\res$. 
  It implies that sterile neutrinos in
the  "broad" resonance (expansion of the Universe is irrelevant)
  equilibrate to a large extent in primordial plasma over
 the typical times $\omega_\res^{-1}$ \eqref{3:sol_answer1}. We verified that
 the interacting rate of such intensity or less does not affect our
  outcomes.}
\begin{equation}\label{fN_low}
\frac{f_N}{f_{FD}}\simeq 1\,.
\end{equation}
%To trace the correspondent dependency of sterile neutrino distribution function we rely on rather general assumptions.
In the opposite limit, $n\sqrt{2}m_\phi H/\omega_\res^2\gg 1$, the
generation of sterile neutrinos is strongly suppressed. The production
rate in the resonance is given by the frequency of corresponding
solution $\omega_\res$, see \eqref{3:sol_answer1}. However, the
"narrow" resonance terminates earlier due to the strong expansion
effect over $\delta t_n\ll \omega_\res^{-1}$ \eqref{res_dur}. Thereby,
the distribution function of the generated during one "narrow" resonance sterile
neutrinos is given
by $f_N/f_{FD}\simeq \omega_\res\delta t_n$ or \eqref{res_dur}
\begin{equation}\label{fN_high}
\frac{f_N}{f_{FD}}\simeq \l\frac{\sqrt{2}m_\phi H}{\omega_\res^2}n\r^{-1}
\end{equation}

%In what follows we focus at the resonance with the highest momentum, that is $n=1$.
%Propagation through the resonance might be ineffective owing to expansion the Universe. “Narrow” resonance is parametrized by \eqref{res_slow}, see Sec.~\ref{subsec:width}. To describe this effect properly 
To find solution interpolating between these two limits \eqref{fN_low}
and \eqref{fN_high} we solve matrix-valued generalizations of the
Boltzmann equations called density matrix equations 
%Flavor
%oscillations, thermal modifications of dispersion relations, and real
%scatterings responsible for de-coherence, can be taken into account
%within one approach
\cite{Sigl:1992fn}
\begin{equation}\label{3:qBoltzman}
i\frac{\partial}{\partial t} \rho =
[\mathcal{H}, \rho]-\frac{i}{2} \{ \Gamma_A, \rho-\rho_\mathrm{eq}\}
\end{equation}
where $\rho$ is the 2x2 density matrix corresponding to active and
sterile neutrinos, with $\rho_{11}$ ($\rho_{22}$) being the
probability density of the active (sterile) neutrino.  Here
$\rho_\mathrm{eq}=\mathrm{diag}(f_\mathrm{FD}(y), f_\mathrm{FD}(y))$
is the equilibrium Fermi-Dirac distribution and
$\Gamma_A=\left( \begin{array}{cc} \Gamma & 0 \\ 0 &
  0\end{array}\right)$ is the damping term due to active neutrino
  interactions in the thermal bath. The evolution starts at early
  times with $\rho=\mathrm{diag}(f_\mathrm{FD}(y), 0)$ and the
interesting  distribution of the produced sterile neutrinos is
obtained as $f_N(y)=\rho_{22}(y)$ at late times, $t\to\infty$. 
We exploit the Hamiltonian in
  vacuum \eqref{3:qHamilton} since its thermal modification brings
  only sub-dominant contribution in accordance with
  Sec. \ref{subsec:plasma}. We also employ precise oscillation
  parameters \eqref{3:transf_matr_thet}, \eqref{3:osc_rate0},
  \eqref{3:transf_matr_m}.

We work in the limit $\omega_\res\gg H$, so the redshift effect in \eqref{3:qBoltzman} can be addressed by the following linear expansion
\begin{align}\label{3:linapp}
\begin{split}
p&=\tilde{p}(1-Ht)\\
M_A&=\tilde{M}_A(1-\frac{3}{2}Ht)
\end{split}
\end{align}
where $\tilde{p}$ and $\tilde{M}_A$ refer to the moment of resonance \eqref{k_def}.

\begin{figure}[!b]
	\centerline{
		\includegraphics[width=0.7\linewidth]{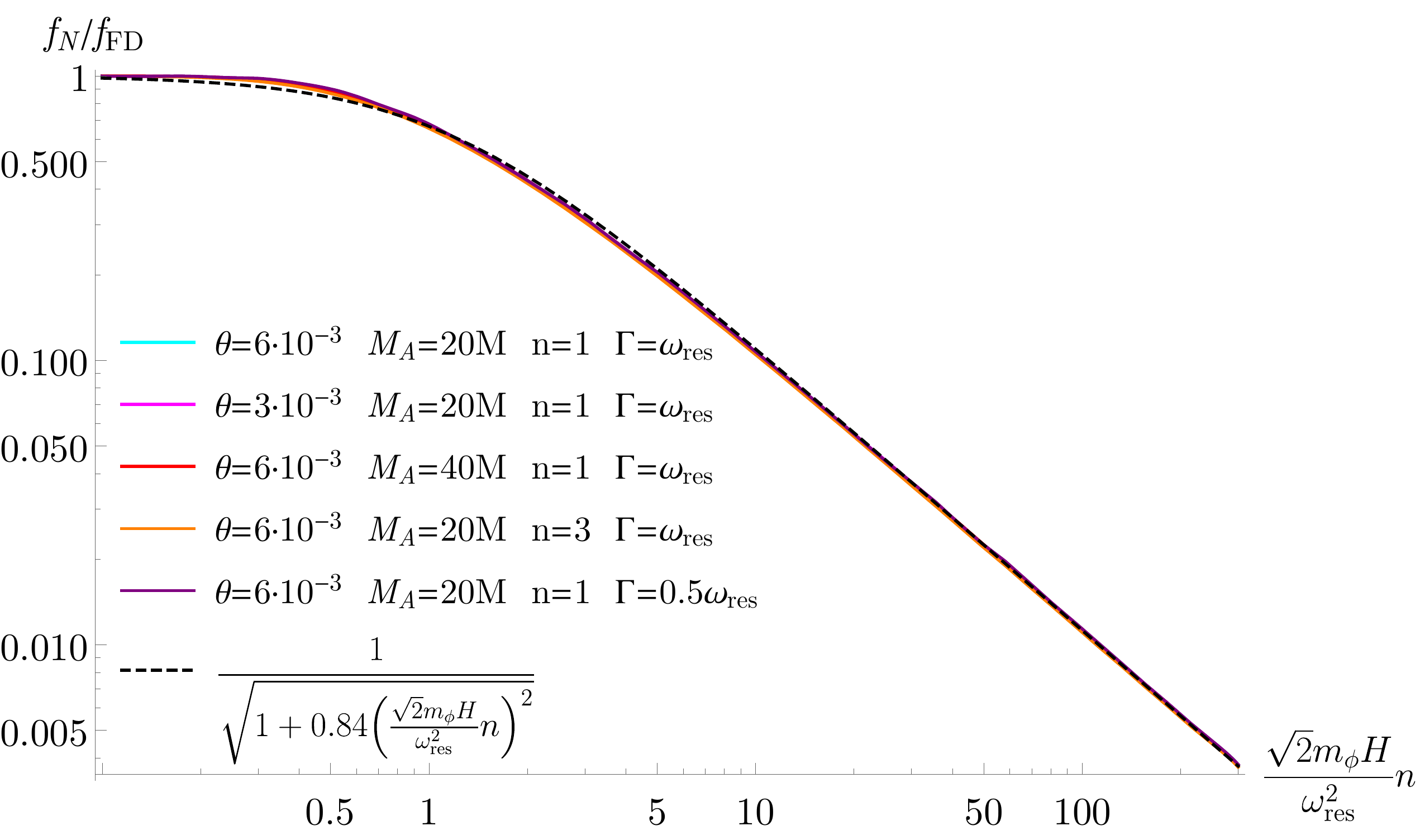}
	}
	\caption{$f_\mathrm{FD}/f_N$ as a function of l.h.s. of \eqref{res_slow}. Dashed black line refers to our fitting result \eqref{3:spect_mH0}.}
	\label{fig:spect_num}
\end{figure}

Solving \eqref{3:qBoltzman}, \eqref{3:qHamilton}, \eqref{3:linapp} for
various values of l.h.s. of \eqref{res_slow} yields the sterile
neutrino distribution function. Results of this numerical analysis for
various model parameters are shown in Fig. \ref{fig:spect_num}. 
Matching of $f_N/f_{FD}$ for various values $\theta$, $M_A/M$, $n$ and
$\Gamma$ demonstrates stability and universal dependency of the
solution on l.h.s. of \eqref{res_slow} which parametrizes the "narrow" resonance in the expanding Universe. We also verified that $f_N/f_{FD}$ is insensitive to $m_\phi$ and to absolute value of $M$. 

For a proper usage we fit our numerical result with the following analytical function
\begin{equation}\label{3:spect_mH0}
f_N = \frac{f_\mathrm{FD}}{\sqrt{1+0.84\l\frac{\disp \sqrt{2}m_\phi H}{\disp	\omega_\res^2}n\r^2}}.
\end{equation}
where $\sqrt{2}m_\phi H/\omega_\res^2$ refers to the moment of
resonance \eqref{k_def}. Since the temperature unambiguously
corresponds to a particular resonant conformal momentum via
\eqref{yk}, our result \eqref{3:spect_mH0} can be presented in a more
conventional form \eqref{omega_res} \cite{Bezrukov:2018wvd}
%\footnote{Our result \eqref{spect_y2} at region  $n\sqrt{2}m_\phi H/\omega_\res^2\sim 1$ slightly differs from that in \cite{Bezrukov:2018wvd}. The reason of that has been a more thorough numerical analysis and. Anyway, such difference alters our results insignificantly and our previous results reproduce the correct limit $f_N/f_{FD}\propto (y/y_s)^{5/2}$ at $y/y_s\rightarrow\infty$.} 
\begin{equation}\label{spect_y2}
  f_N(y) = \frac{f_\mathrm{FD}(y)}{\sqrt{1+0.84\left(\frac{\disp y}{\disp
          y_s}\right)^5}}\,.
\end{equation}
where cut-off scale $y_s$ is defined by 
\begin{equation}\label{3:spect_ys0}
y_s=\frac{0.24}{n^{17/15}}\l\frac{\theta}{4.2\cdot10^{-6}}\r^{4/5}\!\!\l\frac{\!M}{\!\keV}\r^2\!\!\l\frac{9.3\MeV}{T_e}\r^{9/5}\!\!\l\frac{\eV}{m_\phi}\r^{3/5}\!\!\l\frac{50}{g_{*,s}}\r^{1/5}\!\!\l\frac{h_s}{h_e}\r^{3/5}.
\end{equation}
The mode, corresponding to $y_s$ passes through the resonance \eqref{k_def} at temperature \eqref{yk}, \eqref{3:spect_ys0}
\begin{equation}\label{3:spect_Ts0}
T_s=\frac{28\MeV}{n^{1/15}}\l\frac{\theta}{4.2\cdot10^{-6}}\r^{2/5}\!\!\l\frac{T_e}{9.3\MeV}\r^{3/5}\!\!\l\frac{m_\phi}{\eV}\r^{1/5}\!\!\l\frac{50}{g_{*,s}}\r^{1/10}\!\!\l\frac{h_e}{h_s}\r^{1/5}
\end{equation}

The net results is given approximately by the sum of \eqref{spect_y2}
over resonances $n=1,2,\dots$. However, one observes from
\eqref{3:spect_ys0} that the higher resonances are relevant at lower
neutrino momentum. The most
prominent contribution both to neutrino abundance and to the average
neutrino velocity comes from the lowest resonance $n=1$. This case is
studied in the next Section.
%. In this case one should require $y_s\ll 3$. The latter implies that the most important contribution  

%%%%%%%%%%%%%%%%%%%%%%%%%%%%%%%%%%%%%%%%%%%%%%%%%%%%%%%%%%%%%%%%%%%%%%%%%%%%%%%%%%%%%%%%

\section{Sterile neutrino dark matter}
\label{Sec:dark_matter}

In this section we consider sterile neutrino DM that is produced in
resonance and has spectrum \eqref{spect_y2}. The correct DM neutrino
abundance, in this case, can be achieved only if the resonant conformal
momentum \eqref{3:spect_ys0} is small enough. 
Since higher
resonances are relevant at lower neutrino momentum
\eqref{3:spect_ys0}, resonance with the highest momentum ($n=1$) gives
the most important contribution both to the neutrino abundance and to
the average neutrino velocity. Given this reason, we neglect any
contribution of higher resonances ($n>1$) and examine sterile neutrino
production in resonance with $n=1$ only.

\begin{figure}[!htb]
	\centerline{
		\includegraphics[width=0.6\linewidth]{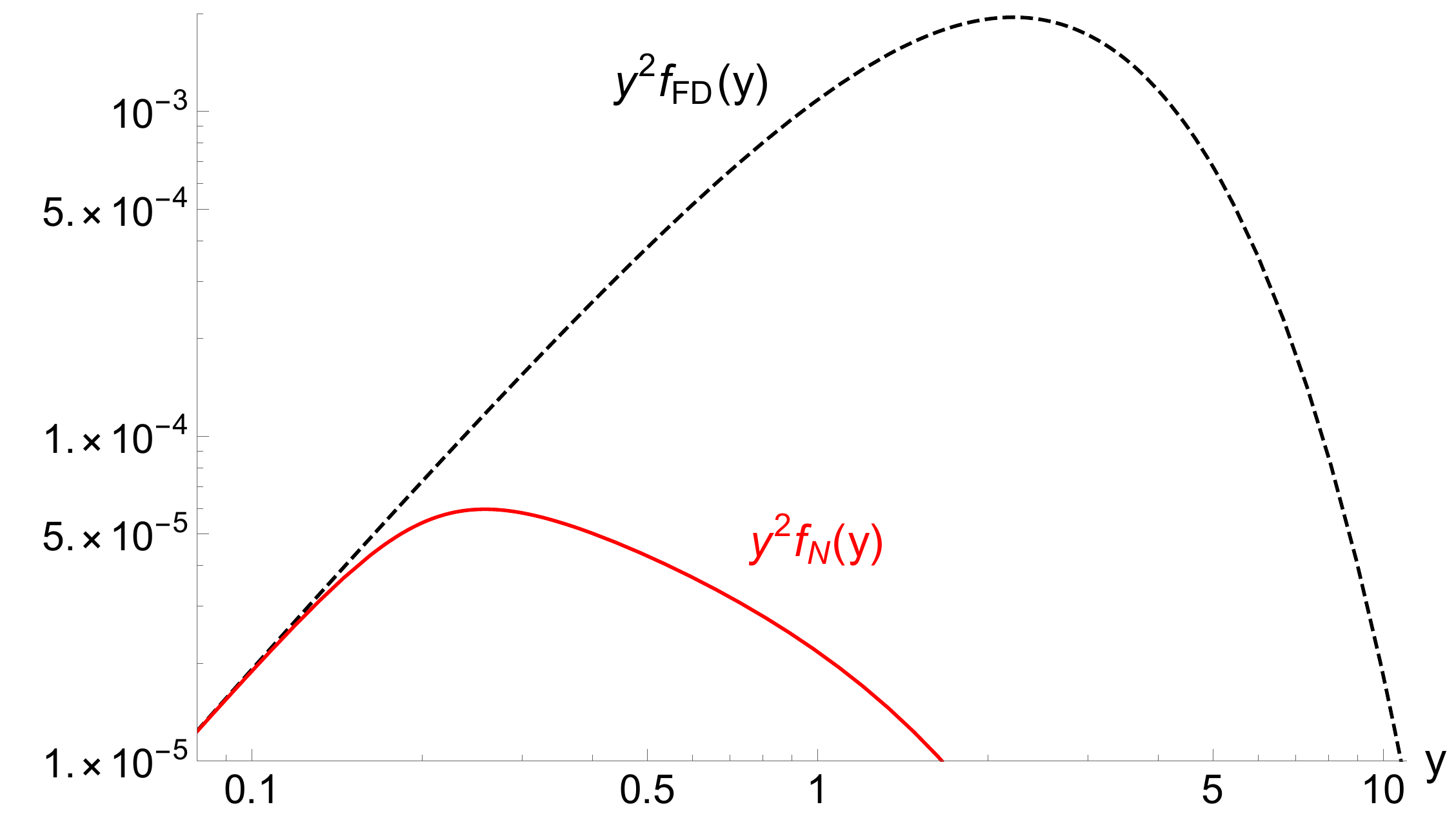}
	}
	\caption{The Fermi--Dirac $f_\mathrm{FD}$ and sterile neutrino $f_N$ spectra in logarithmic scale.}
	\label{fig:spect}
\end{figure}

The proper abundance of DM composed of sterile neutrinos of spectrum~\eqref{spect_y2} is achieved for~\cite{Bezrukov:2018wvd}
\begin{equation}
\label{eq:yg2}
y_s\simeq 0.24\times\left(\frac{1\keV}{M}\right)^{2/5}.
\end{equation}
The sterile neutrino distribution function \eqref{spect_y2} for this cutoff is shown in Fig. \ref{fig:spect}.
The characteristic feature in the sterile neutrino distribution
function at $y\gtrsim y_s$ provides a cool spectrum with average 
\begin{equation}\label{y_aver}
\lAngle y\rAngle=0.6\l\frac{50}{g_{*,s}}\r^{1/3}
\end{equation}
where $g_{*,s}$ refers to the plasma effective degrees of freedom at
the reference temperature of sterile neutrino production $T\simeq
T_s$. Such a feature allows to alleviate structure formation
constraints on the mass of DM particle. Ly-$\alpha$ constraint $m_{\rm
  NRP}>8\keV$ \cite{Adhikari:2016bei} is translated in our case to
\begin{equation}
M>1.5\keV \l\frac{50}{g_{*,s}}\r^{1/3}.
\end{equation}
%If the sterile neutrinos are produced at temperatures $T_s\gg
%100\MeV$, the corresponding constraint is significantly weaker.

\subsection{Parameter constraints}
\label{subsec:constraints}

Here we examine the available parameter region $(\theta^2,M)$ where
the sterile neutrinos with spectrum \eqref{spect_y2}, \eqref{y_aver}
is capable of composing all the present dark matter. For that, we should address all relevant constraints.

For the purpose of this Section one can equate \eqref{eq:yg2} and \eqref{3:spect_ys0} that leads to
\begin{equation}\label{3:spect_theta}
\theta \sim 4.2\times10^{-6}
\left(\frac{1\keV}{M}\right)^{3}
\left(\frac{g_{*,s}}{50}\right)^{1/4}
\left(\frac{T_e}{9.3\MeV}\right)^{9/4}
\left(\frac{m_\phi}{1\eV}\right)^{3/4}
\l\frac{h_e}{h_s}\r^{3/4}
\end{equation}
Further, we assume that sterile neutrino DM is effectively produced at temperature $T\simeq T_s$ \eqref{yk},  \eqref{eq:yg2}
\begin{equation}\label{T_s}
T_s = 3T_e
\left(\frac{T_e}{9.3\MeV}\right)^{1/2}
\left(\frac{m_\phi}{1\eV}\right)^{1/2}
\left(\frac{1\keV}{M}\right)^{6/5}
\left(\frac{h_{e}}{h_{s}}\right)^{1/2}.
\end{equation}
Since we explore here parameter space $(\theta^2,M)$, it is convenient to rewrite \eqref{T_s} in the following form \eqref{3:spect_theta}
\begin{equation}\label{3:Ts_theta}
T_s=28\MeV\l\frac{\theta}{4.2\cdot 10^{-6}}\r^{2/3}\l\frac{M}{\keV}\r^{4/5}\l\frac{50}{g_{*,s}}\r^{1/6}.
\end{equation}
%Below we shortly list all relevant constraints.

%effective temperature production
One of the most rigorous foundations of our theoretical framework
refers to the assumption of significantly large scalar field values. To
address \eqref{cond1} at the moment of sterile neutrino production we
require $T_e<3T_s$, see \eqref{MA}. In addition, we always assumed
that the active neutrinos equilibrate in the thermal bath, so the parameter
space should meet $1\MeV<T_e$. Finally, $T_e$ is confined within
\begin{equation}\label{Te_cond}
1\MeV<T_e<\frac{T_s}{3}\,.
\end{equation}

To provide small thermal modifications and address \eqref{cond2} we impose \eqref{2:VeffG}, \eqref{MA}, \eqref{eq:yg2}, \eqref{T_s}
\begin{equation}\label{plasma_cond}
\left.5V\frac{4yT}{M_A^2}\right\vert_{\substack{y=y_s \\ T=T_s}}<1
\end{equation}

\iffalse
To respond \eqref{3:beta} at reference scales one should require \eqref{MA}, \eqref{eq:yg2}, \eqref{T_s}
\begin{equation}\label{kinetic_cond}
%\left.\frac{M_A}{y_sT}\right|_{T=T_s}<1
\left.\frac{M_A}{yT}\right\vert_{\substack{y=y_s \\ T=T_s}}<1
\end{equation}
\fi

At $T\simeq T_e$ the regime of high scalar field amplitude
\eqref{3:z_small} terminates and the common scenario of the
non-resonant production via active-sterile oscillations in plasma 
\cite{Dodelson:1993je} is
resumed. Sterile neutrino amount produced at $T< T_e$ should be
relativily small, 
\begin{equation}\label{DW_cond}
\Omega_{N,T<T_e}<\OmegaDM, 
\end{equation}  
where $\Omega_{N,T<T_e}$ denotes the corresponding sterile neutrino
contribution, generated at $T<T_e$ ($M_\eff\approx
M$), for details see \cite{Bezrukov:2017ike}.

To avoid the effective scalar decay to sterile neutrinos we limit the
mass scale as  
\begin{equation}\label{mass_up}
m_\phi<M
\end{equation}
When the mass of scalar field belongs to the interval $0.01\eV\approx
\sqrt{\Delta m_{\rm sol}^2}<m_\phi<M$, the scalar field can decay to
active neutrinos. This moment relates to $H\sim
\Gamma_{\phi\rightarrow\nu\nu}=f^2m_\phi/16\pi$ 
and hence happens when the at the temperature $T\simeq T_\dec$ such
that 
\begin{equation}\label{3:Tdec0}
\frac{T_\dec}{T_0}=\frac{\theta^2}{4\sqrt{\pi}}\,\frac{f}{\Omega_{\rm rad}^{1/4}}\l\frac{m_\phi}{H_0}\r^{1/2}\l\frac{g_{*,0}}{g_{*,\dec}}\r^{1/4}.
\end{equation}
This decay channel does not affect sterile neutrino production at
$T\simeq T_s$ if the corresponding decay is late enough
\eqref{3:Tdec0}, \eqref{T_s}, i.e. 
\begin{equation}\label{Tdec_cond}
%\left.T_\dec\right|_{f=f_{\rm min}}<T_s
T_\dec<T_s
\end{equation}

We suggest that the perturbative treatment is applied to the scalar
field sterile neutrino interaction and hence assume that \eqref{3:fmin} 
\begin{equation}\label{fmin_cond}
f<1\,.
\end{equation} 
It's worth noting that the Yukawa coupling \eqref{eq:LNh} enters
conditions \eqref{Tdec_cond} and \eqref{fmin_cond} only. As we are
interested in the most extended attainable region in model parameter
space, one can fix $f$ in its lower boundary coming from the dark
matter overproduction constraint,  $\Omega_\phi<\Omega_{\rm DM}$, 
\begin{equation}\label{3:fmin}
f_{\rm min}^2=\frac{m_\phi^2M^2}{2\Omega_{DM}\rho_\mathrm{crit}}
\frac{h_{0}T_0^3}{h_{e}T_e^3}.
\end{equation}

Finally, 
sterile neutrino dark matter should be astrophysically viable. That
means, the vacuum mixing  must agree with X-ray constraint
\begin{equation}\label{Xray_cond}
\theta<\theta_{\rm X-ray}(M). 
\end{equation}

\iffalse
We also recall that our estimates are valid only for coherent evolution of neutrino states, so \eqref{2:VeffG}, \eqref{omega_res}
\begin{equation}\label{w_cond}
\left.\frac{\Gamma}{\omega_\res}\right\vert_{\substack{y=y_s \\ T=T_s}}<1
\end{equation}
\fi

\subsection{Results}
\label{subsec:results}

\begin{figure}[!t]
	\centerline{
	\includegraphics[width=0.8\linewidth]{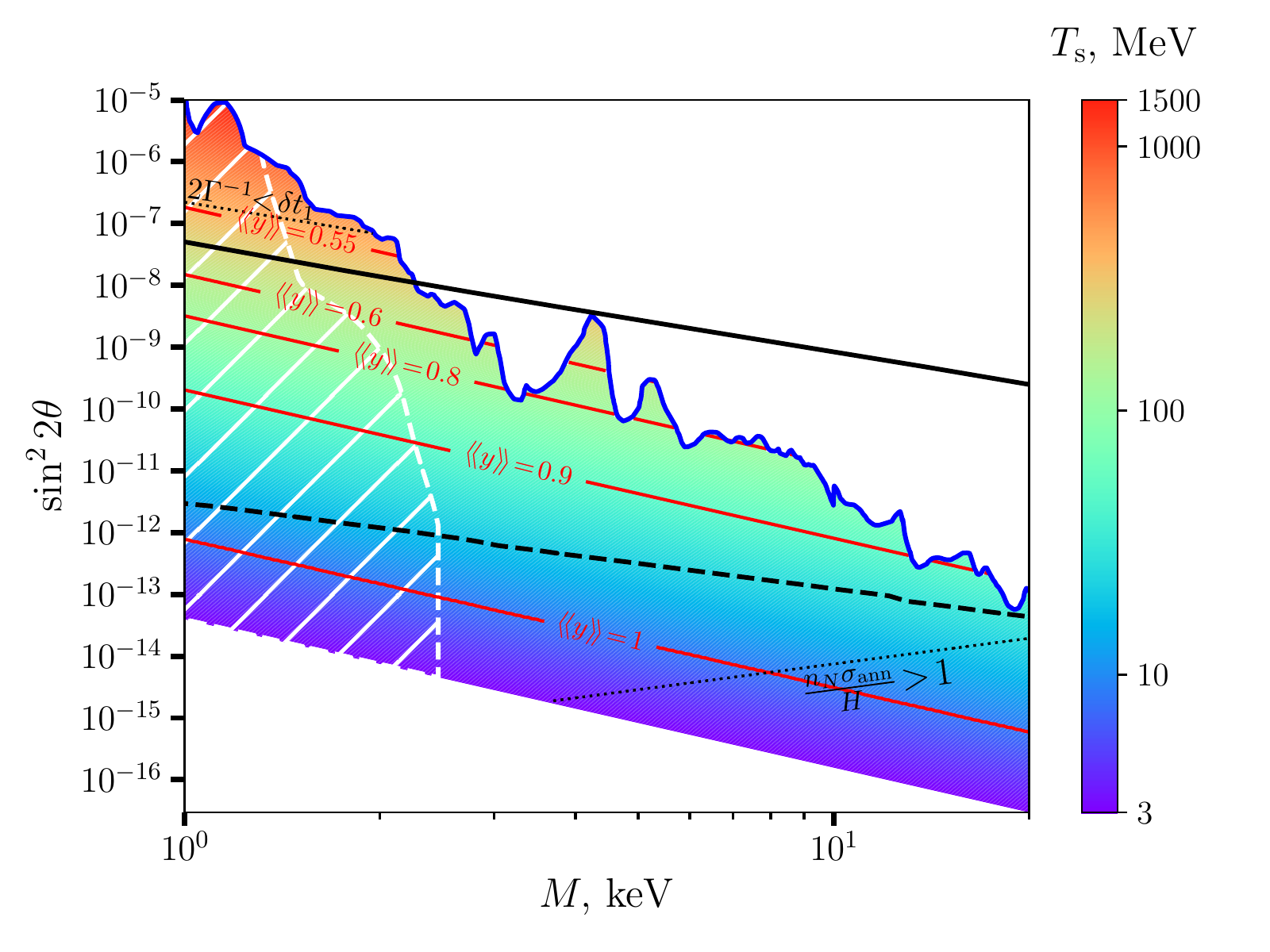}
	%\hspace{-0.8cm}
	%\includegraphics[width=0.6\linewidth]{region2}
	\vspace{-0.5cm}
}
	\caption{Parameter scan over the sterile neutrino mass and
          active-sterile mixing angle (the corresponding scalar masses
		$10^{-2}\eV<m_\phi<M$ help to avoid additional production of
		sterile neutrinos from scalar decays at $T\simeq m_\phi$ as in
		Refs.\    \cite{Shaposhnikov:2006xi,Kusenko:2006rh,Petraki:2007gq}
                and $1\MeV<T_e<150\MeV$). The blue
		line is the upper limit from X-ray
		observations~\cite{Roach:2019ctw,Malyshev:2014xqa,Horiuchi:2013noa}. The colour region gives the
		fraction of the sterile neutrino equal to DM, 
		$\Omega_N=\Omega_{DM}$. The colour indicates the
		reference temperature of the sterile neutrino
                production $T_s$ \eqref{3:Ts_theta}.  The white dashed region is excluded by studying the cosmic structure
		formation. Red lines refers to sterile neutrino average momenta per temperature calculated just before active neutrino freezeout \eqref{y_aver}. For reference: the black line is for the conventional
		non-resonant generation mechanism~\cite{Adhikari:2016bei}, black
		dashed line corresponds to the maximal lepton asymmetry attainable
		in the $\nu$MSM~\cite{Adhikari:2016bei}, upper dotted line relates to equality of \eqref{wCond}, lower dotted line corresponds to equality of  \eqref{3:ann0}.}
	\label{fig:region}
\end{figure}

At each point of parameter space $(\theta^2,M)$ we look over $\langle
T_e,m_\phi\rangle$ and find optimal parameters which address
\eqref{Xray_cond}, \eqref{Te_cond}, \eqref{plasma_cond},
\eqref{DW_cond}, \eqref{mass_up}, \eqref{Tdec_cond} and
\eqref{fmin_cond} satisfactory. As a result of this procedure we find
the most extended available region on the plane ($M,\sin^22\theta$),
which is depicted in Fig. \ref{fig:region}.

% \eqref{w_cond}
The lower boundary on the mixing angle in the plot of
Fig. \ref{fig:region} refers to $T_s=3T_e=3\MeV$, see
\eqref{Te_cond}. At lower values of $\theta$ the temperature $T_e$
drops below $1\MeV$ and with the active neutrino out of equilibrium
our analysis becomes more involved. The maximum possible mixing angle,
in turn, is defined by X-ray constraint \eqref{Xray_cond}. Condition
\eqref{DW_cond} requires $T_e\lesssim 150 \MeV$ above the non-resonant
production line \cite{Bezrukov:2018wvd} which does not affect 
our region in the model parameter plane ($M,\sin^22\theta$): the corresponding constraint on $\theta$ is always weaker
than that from the X-ray in Fig. \ref{fig:region}.
%only for significant large mixing $\sin^22\theta>10^{-5}$ which is beyond the scope of Fig. \ref{fig:region}.
%For large values of $\theta$ which are close to non-resonant production line, there is also a contribution from DW-like active-sterile oscillations at $T<T_e$ provided by \eqref{DW_cond}. This leads to additional constraints on $T_s$ which value is no longer defined by \eqref{3:Ts_theta}. In this case sterile neutrino amount is saturated at $T<T_e$ and obviously can not provide with DM today due to stricture formation constraint on thermal  populated particles.

One comment is in order here. Numerical analysis carried out in Sec. \ref{Sec:spectrum} holds in case of small enough interaction rates, namely $\Gamma<\omega_\res$. However, the resonances which bring the bulk contribution to the sterile neutrino dark matter are narrow, see \eqref{spect_y2}, \eqref{eq:yg2}. It implies that typical time of neutrino conversions in resonance is small compared to $\omega_\res^{-1}$ and the corresponding constraint on $\Gamma$ is weaker. All in all, the interaction rate does not affect the oscillations in resonance if the interaction length in plasma surpasses the typic time of resonant oscillations \eqref{res_dur}, i.e.
\begin{equation}\label{wCond}
2\Gamma^{-1}>\delta t_1
\end{equation} 

We outline the region where condition \eqref{wCond} is broken by dotted line in the top part of Fig. \ref{fig:region}. The parameter space above this line cannot be reliably described by outcomes of Sec. \ref{Sec:spectrum} and needs a more sophisticated approach which accounts for interplay between flavor oscillations and de-coherence processes induced by real scattering. We verified that the interaction rate of large intensity efficiently suppresses sterile neutrino production in resonance. Hence, assuming inefficient non-resonant generation one can easily produce a sterile neutrino dark matter in the regime of strong interaction $2\Gamma^{-1}<\delta t_1$. However, in this parameter space momentum distribution of dark matter particles is no longer described by \eqref{spect_y2} and one needs a more proper solving of \eqref{3:qBoltzman} which is beyond the scope of this paper. Since the regime of strong interaction refers to significantly small sterile neutrino masses where structure formation constraints become crucial the validity of the region with relatively large mixing in top of Fig. \ref{fig:region} inquires a further investigation. To be more conservative in what follows we will assume $\lAngle y\rAngle>0.55$ in accord with Fig. \ref{fig:region}. We emphasize that the scalar induced resonance can lead to substantially cooler velocity distribution of sterile neutrinos as compared to \cite{Bezrukov:2018wvd}. The reason of this is the refined treatment of resonant oscillations in the high temperature regime provided in Sec. \ref{subsec:plasma} which was previously unexplored.

%flavor oscillations and scattering processes responsible for de-coherence of quantum state.
%Interesting result of provided analysis consists in that the mass of scalar field is restricted in the following interval $10^{-2}\eV<m_\phi<M$. It precludes the scalar decay to sterile neutrinos considered in Refs.\    \cite{Shaposhnikov:2006xi,Kusenko:2006rh,Petraki:2007gq}

To provide a straightforward comparison of sterile neutrino velocity distribution among different generation mechanism of sterile neutrino dark matter we list average momenta per temperature in various scenarios calculated just before active neutrino freeze-out corresponding to $g_*= 10.75$ in Tab. \ref{table:comp}.    
\begin{table}[h!]
	\centering
	\begin{tabular}{|l|c|c|}
		\hline
		Model &	$\lAngle p/T\rAngle$	& References \\
		\hline
		Dodelson-Widrow	& 2.8 & \cite{Dodelson:1993je,Laine:2008pg} \\
		Shi-Fuller	& 1-2.8 & \cite{Shi:1998km,Laine:2008pg}\\
		Scalar decay at $T\sim100\GeV$  	& 1.1 & \cite{Shaposhnikov:2006xi,Petraki:2007gq}\\
		Entropy production in dark sector &	0.3-1 & \cite{Bezrukov:2009th,Kusenko:2010ik}\\
		Scalar induce resonance  	& 0.55-1 & this work\\
		\hline
	\end{tabular}
	\caption{\label{table:comp} A summary of the models with their respective average momenta per temperature rescaled to the moment before active neutrino freezeout corresponding to $g_*= 10.75$ with relevant references.}
\end{table}
Scalar induced resonance provides the coldest sterile neutrino distribution among all known mechanisms relying on active-sterile mixing, except for those who have significant entropy release after the production of the sterile neutrino like Grand Unified Theory \cite{Kusenko:2010ik}. So, the proposed mechanism opens a wide area of sterile neutrino masses compatible with dark matter.  Discovery of sterile neutrino dark matter with low masses would thus favour the current mechanism.  Moreover, as far as non-vanishing active-sterile mixing
angle is required in the model, the observation of X-ray signal from the dark matter decay is predicted and potentially testable by the future X-ray satellite missions.
%We emphasize that quantities provided in Tab. \ref{table:comp} should be rescaled to compare with results of \cite{Abazajian:2019ejt}

%%%%%%%%%%%%%%%%%%%%%%%%%%%%%%%%%%%%%%%%%%%%%%%%%%%%%%%%%%%%%%%%%%%%%%%%%%%%%%%%%%%

\section{Other issues}
\label{Sec:issues}

The amplification of sterile neutrino production in the early Universe
we discuss in this paper is based on a rather general observation that
a non-standard cosmological evolution of the model parameters can have
drastic consequences for the oscillating active-sterile neutrino
system. We have demonstrated this by introducing (possibly) the
simplest ingredient --- a massive light scalar coupled to the sterile
neutrino. This choice is adhoc but plays its role nicely, illustrating
many important features of the suggested mechanism. However, the model
with only one additional scalar field is not realistic.  A fully
realistic extension of SM (which has, for example, more fields and
interactions) one has to solve the set of typical problems, unrelated
to the main idea of the mechanism. In this section we intend to
consider those problems and its possible solutions.

%\subsection{Extended scalar sector}

In a realistic model the scalar potential can be naturally more
complicated: both in form and in content.

To illustrate the point, take the simple scalar model we used
\eqref{eq:LNh}, \eqref{eq:phi}. The scalar is not free, it couples to
sterile neutrinos, and hence generically induces quantum corrections
introducing more interaction terms in the model lagrangian. Naturally,
one expects additional $\phi^2$, $\phi^3$ and $\phi^4$ terms. The
first term is renormalization of the scalar mass, which alike the SM
Higgs mass is not protected and suffers from the quadratic
divergences. The second term originates from logarithmically divergent
Feynman diagrams, and it is proportional to the sterile neutrino mass
reflecting remnants of the lepton symmetry in the model. The third
term is of order $y^4f^4/(64\pi^2)$ and also comes from logarithmically divergent diagrams.

All the three terms are dangerous for the scalar vacuum because they
tend to destabilize it. Indeed, the cubic $\phi^3$ term delivers
explicitly opposite sign contributions at $\phi\to\pm\infty$. The
quartic $\phi^4$ term is negative and grow in value with the energy
scale, and the scalar quantum potential becomes negative at large
fields exceeding a certain finite normalized value. Finally, a naive
estimate of the divergent contribution with momentum cutoff reveals
the negative mass squared for the quadratic $\phi^2$ term, which may
be meaningless given the major hierarchy problem for the scalar mass
scale $m$. A specific mechanism can be invoked to cancel the quantum
corrections, like those (e.g., supersymmetry, technicolor) introduced
to protect the electroweak scale and cure the gauge hierarchy problem
for the SM Higgs.

While the vacuum stability is an important issue for the theory
itself, the new terms in the scalar potential may well participate in
the scalar field dynamics of the expanding Universe. In particular,
the $\phi^4$ term dominating over $\phi^2$ term changes the
time-dependence of the oscillating field amplitude, and hence the
time-dependence of the sterile neutrino mass. The active-sterile
neutrino system still exhibits resonant behaviour in this case, but
the numerical results differ. Apart of that, if $\phi^4$ term
dominates in the late Universe as well, the scalar contributes to the
dark radiation component, rather than to the dark matter. One has to take
care of the potential scalar impact on the cosmology, with a possible
interplay between $\phi^4$- and $\phi^2$-dominating regimes. In
specific situations the impact may be negligible, e.g., with scalar
potential vanishing in the late Universe. The scalar forms stiff
matter and its energy density disappears (no relevant limits on
$\Omega_\phi$ then).

Generally, with several degrees of freedom in the scalar sector, the
situation becomes more complicated. Some of them may be involved into
cosmological dynamics and impact on both the Universe evolution and
the sterile neutrino production process. The natural physically
motivated extension of our model includes Majoron --- the Goldstone
scalar emerging after the spontaneous breaking of the lepton
symmetry. This framework implies promotion of our scalar $\phi$ to
the complex scalar, which complex phase is associated with the
Majoron. It is massless, and hence may contribute to the dark
radiation of the Universe (if inhomogeneous) or form stiff
matter (kinetic term dominates).
Another example is familon in the extensions where the flavor
structure of the neutrino sector is developed. 

Apart of different homogeneous evolution at the time of sterile
neutrino production, new degrees of freedom may change the late time
behaviour of the system, e.g.\ by inducing the scalar $\phi$ decay
(hence invaliding limits on $\Omega_\phi$). Coupling to other fields
may contribute to the scalar mass, so that $\phi$ becomes heavy in the
late Universe. In this way the scalar decay rate can increase, and
even plasma processes may change. Indeed, within our simple model, in
the late-time Universe the 
%\subsection{Sterile neutrino annihilation}
annihilation processes $NN\rightarrow\phi\phi$ can dilute sterile neutrino abundance produced in resonance. Using number density of sterile neutrinos  $n_{N}=2\frac{4}{11}T^3\int4\pi y^2f_N(y)dy$ \eqref{spect_y2}, \eqref{eq:yg2}, characteristic annihilation cross section $\sigma_{\rm ann}\sim10^{-2}f^4/T^2$ and Hubble $H$ efficiency of such a process at $T=M$ should be small \eqref{3:fmin}
\begin{equation}\label{3:ann0}
\left.\frac{n_{N}\sigma_{\rm ann}}{H(T)}\right\vert_{T= M}\approx0.8\l\frac{f_{\rm min}}{4\cdot 10^{-4}}\r^4\l\frac{\keV}{M}\r\l\frac{g_{*,0}}{g_{\rm *,ann}}\r^{1/2}<1
\end{equation}
This constraint and $T_e<T_s/3$ \eqref{Te_cond} induce a new lower
boundary for relativily large $M$ outlined by dotted line in bottom of
Fig. \ref{fig:region}. 
With much heavier scalars in the late Universe the annihilation is
kinematically forbidden. Otherwise one must generate more sterile
neutrinos to have $\Omega_N=\Omega_{DM}$ and take care of the decay
products of the scalars in the late Universe.    

Likewise, the sterile neutrino component itself may impact on the
scalar field via back reaction process. Even if the source of sterile
neutrinos is active neutrinos in plasma (not the scalar field itself,
which is also one of the options investigated in
Ref.\,\cite{Bezrukov:2018wvd}) their population may change the
effective potential of the scalar in primordial plasma of the
expanding Universe (and so may do other components in a realistic
extension of the SM). One can estimate the possible effect by Yukawa
coupling as
\[
y\bar N^cN\phi + \text{h.c.} \;\; \to \;\; yn_N\phi\,,
\]
where $n_N$ is the sterile neutrino number density. This term
contributes to the equation of motion for the scalar field by inducing
the external force. The latter shifts $\phi$ from its minimum at
zero to the new value around which the scalar oscillates. Naturally
the scalar contribution to the sterile neutrino mass changes as well
influencing the system dynamics. This procedure induces $y^2 (\bar
N^cN)^2/m_\phi^2$ term in the sterile neutrino sector, which gives the
effective potential suppressing the sterile neutrino production in
plasma, when the density reaches the value high enough for the
induced potential to balance the mass term  
\[
\frac{y^2}{m_\phi^2}n_N\sim M_N\sim y\phi 
\]
This implies the relation
\[
M_Nn_N \sim \frac{M_N^2m_\phi^2}{y^2}\sim m_\phi^2\phi^2\,,
\]
that is equality between the neutrino and scalar energy
densities. Since both quantities equally degrade in the expanding
Universe, the back reaction prevents the sterile neutrino component to
dominate over the scalar one in the late Universe, if the scalar
remains the same as in our minimal model. New degrees of freedom and 
new interaction terms may change this situation (e.g.\ the scalar field
may disappear in the late Universe in the situations we mentioned
above).

\iffalse
\begin{figure}[!t]
	\centerline{
		\includegraphics[width=0.6\linewidth]{region}
		\includegraphics[width=0.6\linewidth]{region2}
	}
	\caption{See Fig. \ref{fig:region}.}
	\label{fig:region2}
\end{figure}
\fi

%\subsection{Quantum corrections}

\section{Summary and prospects}
\label{Sec:sum}

We explored the parametric resonance phenomenon in active-sterile
oscillations with oscillating cosmic background coupled to the
sterile state. Considering the massive light scalar field for
illustrative purposes we developed a new theoretical framework which
yields the time evolution of neutrino probability function at and
near the resonance point. Apart of this, our analytical pipeline
allows one to systematically address several effects such as flavour
oscillations, thermal modifications and expansion of the Universe
which make our approach applicable in cosmology.

Using the elaborated framework we showed that the parametric resonance induced
by the oscillating scalar field can be responsible for the sterile
neutrino dark matter
production in the early Universe. The designed mechanism has several
advantages over other generation scenarios commonly discussed in the
literature. First, it provides the coldest relic sterile neutrino momentum
distribution compared to all other mechanisms relying on
active-sterile oscillations (except models with entropy production in
the dark sector, see discussion at the end of
Sec. \ref{Sec:dark_matter}). Thus, the scalar induced resonance opens
a window of lower mass dark matter, which is otherwise forbidden by
strong constraints from the cosmic structure formation. Second, this mechanism
operates even for very small mixing angle with active neutrinos, thus
evading the X-ray constraints. Therefore, further searches for the
peak-line signatures with the new generation of X-ray telescopes,
e.g. eRosita and ART-XC are justified \cite{Pavlinsky:2008ecy,Merloni:2012uf}. 
Third, the oscillating background itself
can be a source of particles. Sterile neutrinos directly produced by
the scalar field are completely non-relativistic, and hence 
avoid any structure formation constraints, see
\cite{Bezrukov:2018wvd}.
% Finally, the light scalar field appears in various motivated extensions of the SM:
% models with axion-like particles, etc.

% \todo[inline]{FB: However, in models with axion the required coupling is not present. Or you can point me to an example when it is?}

The present manuscript improves the previous analysis \cite{Bezrukov:2018wvd} in several aspects. First, we generalise the analytical framework in the presence of matter, see Sec. \ref{subsec:plasma}. This extension allows us to investigate sterile neutrino production at a higher temperature which results in substantially cooler spectrum with average neutrino momentum down to $p=0.55\cdot T$ compared to $p=T$ as in Ref. \cite{Bezrukov:2018wvd}. This improvement is crucial in the light of strong structure formation constraints. Moreover, the refined treatment makes our mechanism competitive with other popular scenarios of sterile neutrino production relying on active-sterile mixing, see Tab. \ref{table:comp}. Second, we discussed various issues related to the direct implementation of the developed mechanism in full realistic situations. In Sec.~\ref{Sec:issues} we list the essential parts of any realistic model (quantum corrections, subsequent sterile neutrino annihilation, backreaction, etc) and assess their influence on final outcomes. We also outline several natural physically motivated extensions capable of addressing these issues. These findings pave the way to construct a fully realistic and self-consistent extension of SM capable to generate sterile neutrino dark matter.

We stress that the application of our analytical pipeline is not
limited by the case considered in this paper. Our theoretical approach
can be applied to general periodically varying scalar field coupled to
the massive sterile state. For instance, our analysis can be extended
by adding $\phi^4$ term which is motivated by quantum corrections, see
Sec. \ref{Sec:issues}. Possible interplay between
$\phi^2$-$\phi^4$-dominating regimes makes resonance dynamics more
involved with intriguing outcomes for cosmology. The novel analytical
approach designed in this paper can be straightforwardly adopted to describe the
parametric resonance phenomenon in other physical situations with
several oscillators and periodic external fields involved.
However, in a realistic model the backreaction of the produced
particle has to be taken into account, which can further change the results of the analysis.
%In
%particular, our method can be applied to examine the resonance
%dynamics of spin-flavour oscillation in the presence of periodically
%varying electromagnetic fields. Our framework mainly relies on the
%assumption of large scalar field \eqref{3:beta_m} which to our
%knowledge has never been used before in this field. Given this reason
%the elaborated framework is complementary to ones
%\cite{Egorov:1999ah,Dvornikov:2000mt,Dvornikov:2004ms} and can be used
%to examine the resonance dynamics in different assumption about the
%external field.

\begin{acknowledgments}
	The study of resonant production (AC and DG) is supported by RSF
	grant 17-12-01547. The work of FB is supported in part by the
	Lancaster-Manchester-Sheffield Consortium for Fundamental Physics,
	under STFC research grant ST/L000520/1
\end{acknowledgments}

\appendix

\section{Validity of assumptions}
\label{Sec:second_order}

The theoretical approach of Sec. \ref{Sec:resonance}
based on the stationary phase method relies on several
assumptions. Two of them \eqref{3:adiab_constr0}, \eqref{angle_constr}
allow one to reduce rather complicated framework \eqref{Schrod_mass_bas}
to much more primitive form \eqref{ampl_eq_syst}. Here we examine the
validity of this transition.

%As was mentioned in Sec.~\ref{Sec:second_order}, assuming \eqref{3:ampl_eq0} the last term in \eqref{ampl_eq} can be neglected \footnotemark[\getrefnumber{last_term}]. Dependency on $\theta$ in this case seems to be faded, however it remains at initial conditions of \eqref{ampl_eq}. First, we will solve \eqref{ampl_eq} without the last term. Second, we reveal the role of neglected term and construct the appropriate solution.
%Subsequent evolution of $|y_2(t)|$ in this case seems to be totally independent of the mixing~\footnote{Dependency on $\theta_0$ remains for initial conditions of~\eqref{ampl_eq}} and 

Assumptions \eqref{3:adiab_constr0} and \eqref{angle_constr} restrict
the mass scale \eqref{Meff} which can be addressed by
\eqref{ampl_eq_syst}. One of them \eqref{angle_constr} implies
$|\!\sin\theta_\eff|\ll1/2$ \eqref{3:angle_of_time} which can be
recast by making use of \eqref{Meff}, \eqref{neutrino-mass-fall}, \eqref{3:osc_rate} to
\begin{equation}\label{3:theta_constr}
|z+\sin m_\phi t|\gg 2\theta z\,.
\end{equation}
The adiabaticity condition \eqref{3:adiab_constr0}, in turn, can be
transformed to \eqref{3:osc_rate0}, \eqref{3:angle_of_time}, 
\begin{equation}\label{3:adiab_constr}
|z+\sin m_\phi t|\gg\l\theta z\frac{m_\phi}{2\beta}\r^{1/4}.
\end{equation}

%We need one more mass scale related to resonance itself that can be
%confronted with either \eqref{3:theta_constr} or
%\eqref{3:adiab_constr}.
To understand where the resonance feature
manifests itself we scrutinize the neutrino system evolution during one
oscillating period of the scalar field. Evident transformations help 
to reformulate \eqref{ampl_eq_syst} in the following second-order
differential equation on the transition probability $y_2(t)$, 
\begin{equation}\label{ampl_eq}
\frac{\partial^2y_2(t)}{\partial t^2}-\frac{\partial y_2(t)}{\partial t}\left[\!2i\beta(z+\sin m_\phi t)^2+\frac{m\cos m_\phi t}{z+\sin m_\phi t}\!\right]+4y_2(t)\beta^2z^2\theta^2(z+\sin m_\phi t)^2=0,
\end{equation}
with inital conditions $y_2(0)=0$, $\frac{\disp\partial y_2}{\disp\partial t}(0)=-2i\beta z^2\theta$, see~\eqref{ampl_eq_syst}.

The Sturm--Liouville theory~\cite{Romanko} allows to present the solution of~\eqref{ampl_eq} in the following form
\begin{equation}\label{y_comp}
y_2(t)=u(t)\cdot w(t)\,,
\end{equation}
where $u(t)$ can be found analytically:  
\begin{equation}
u(t)=e^{\half\int_0^t\left[2i\beta(z+\sin \zeta)^2+m_\phi\cos m_\phi\zeta/(z+\sin m_\phi\zeta)\right]d\zeta}=\sqrt{\left|\frac{z+\sin m_\phi t}{z}\right|}e^{i\gamma},
\end{equation}
with $\gamma=\int_0^t\beta(z+\sin m_\phi\zeta)^2d\zeta$~\footnote{Since we are interested in probability amplitude $|y_2(t)|$, we will often neglect a pure phase multiplier in what follows.} and $w(t)$ obeying
\begin{align}\label{ampl_eq_SL}
&\frac{\partial^2w(t)}{\partial t^2}+q(t)\cdot w(t)=0,\\
\label{ampl_eq_SLfr}
\begin{split}
q(t)\equiv\beta^2(z+\sin m_\phi t)^4-\frac{3}{4}&\frac{m_\phi^2\cos^2m_\phi t}{(z+\sin m_\phi t)^2}+i\beta m_\phi(z+\sin m_\phi t)\cos m_\phi t\\
&-\half\frac{m_\phi^2\sin m_\phi t}{z+\sin m_\phi t}+4\beta^2z^2\theta^2(z+\sin m_\phi t)^2.
\end{split}
\end{align}

To begin with, we note that our assumption \eqref{3:theta_constr}
implies the hierarchy 
\begin{equation}\label{3:ampl_eq0}
4\beta^2z^2\theta^2(z+\sin m_\phi t)^2\ll\beta^2(z+\sin m_\phi t)^4\,.
\end{equation} 
It means that we can neglect the term $4\beta^2z^2\theta^2(z+\sin
m_\phi t)^2$ in~\eqref{ampl_eq_SLfr}, since it gives a subdominant
contribution to $q(t)$ which governs dynamics of the system
\eqref{ampl_eq_SL}. 
%\footnote{\label{last_term} In fact, term $4\beta^2z^2\theta^2(z+\sin m_\phi t)^2$ can be important if long evolution times are considered. The role of this term will be revealed later on.
	%In fact, term $4\beta^2z^2\theta^2(z+\sin m_\phi t)^2$ is responsible for backreaction in two-level system. Thus, our consideration implies $|y_2(t)|\ll1$. For more details see discussion on this topic in the beginning of Sec.~\ref{Sec:phase_method}.
	%}.
Now, let us consider two limits: $\l\frac{\disp
  m_\phi}{\disp\beta}\r^{1/3}\ll|z+\sin m_\phi t|\leq1$ and $|z+\sin
m_\phi t|\ll\l\frac{\disp m_\phi}{\disp\beta}\r^{1/3}$. In the first
case we obtain \eqref{3:beta_m}: 
\begin{align}\label{3:ampl_eq1}
\begin{split}
&\frac{3}{4}\frac{m_\phi^2\cos^2m_\phi t}{(z+\sin m_\phi t)^2}\ll\beta^2(z+\sin m_\phi t)^4,\\
&|\beta m_\phi(z+\sin m_\phi t)\cos m_\phi t|\ll\beta^2(z+\sin m_\phi t)^4,\\
&\left|\half\frac{m_\phi^2\sin m_\phi t}{z+\sin m_\phi t}\right|\ll\beta^2(z+\sin m_\phi t)^4.
\end{split}
\end{align} 
In the second case one gets \eqref{3:beta_m}, \eqref{3:z_small}: 
\begin{align}\label{3:ampl_eq2}
\begin{split}
&\beta^2(z+\sin m_\phi t)^4\ll\frac{3}{4}\frac{m_\phi^2\cos^2m_\phi t}{(z+\sin m_\phi t)^2},\\
&|\beta m_\phi(z+\sin m_\phi t)\cos m_\phi t|\ll\frac{3}{4}\frac{m_\phi^2\cos^2m_\phi t}{(z+\sin m_\phi t)^2},\\
&\left|\half\frac{m_\phi^2\sin m_\phi t}{z+\sin m_\phi t}\right|\ll\frac{3}{4}\frac{m_\phi^2\cos^2m_\phi t}{(z+\sin m_\phi t)^2}.
\end{split}
\end{align}

Then, according to \eqref{3:ampl_eq0}, \eqref{3:ampl_eq1} and
\eqref{3:ampl_eq2}, the asymptotics for $q(t)$ in \eqref{ampl_eq_SLfr} equal 
\begin{align}
%\begin{split}
\label{osc_vs_growth1}
|z+\sin m_\phi t|&\gg\l\frac{\disp m_\phi}{\disp\beta}\r^{1/3}\rightarrow q(t)\approx\beta^2(z+\sin m_\phi t)^4>0\,,\,\,\,\,\,\, \\
\label{osc_vs_growth2}
|z+\sin m_\phi t|&\ll\l\frac{\disp m_\phi}{\disp\beta}\r^{1/3}\rightarrow q(t)\approx-\frac{3}{4}\frac{m^2\cos^2m_\phi t}{(z+\sin m_\phi t)^2}<0\,.\,\, 
%\end{split}
\end{align}
Since the squared frequencies in the two considered limits are of
different signs, we refer to \eqref{osc_vs_growth1} and
\eqref{osc_vs_growth2} as oscillations and dumping regimes, respectively.

%Here, only the first term in \eqref{ampl_eq_SLfr} brings dominant contributions to $q(t)$. Since this contribution is positive we refer such regime as oscillating.
%Thus, only the second term in \eqref{ampl_eq_SLfr} brings dominant negative contributions to $q(t)$. For the reason that would be clarified latter we refer such regime as dumping.

We start with oscillation regime \eqref{osc_vs_growth1}. We examine
solution of \eqref{ampl_eq_SL} near the point $z+\sin m_\phi t_0= 1$
and obtain $w(t)=C_1\sin\l\beta t+C_2\r$. Assuming $w(t_0)=0$,
$\partial w(t_0)/\partial t=-2i\beta z\sqrt{z}\theta$, see
\eqref{y_comp} and \eqref{ampl_eq_syst}, we arrive at
\begin{equation}\label{y2_osc}
y_2(t)=-2iz\theta\sin\left[\beta \l t-t_0\r\right]. 
\end{equation}
Solution \eqref{y2_osc} in the vicinity of $z+\sin m_\phi t_0=1$ describes oscillations with regular amplitude $2z\theta$.

In the regime \eqref{osc_vs_growth1} we explore the solution of \eqref{ampl_eq_SL} near the point $\!z+\sin m_\phi t_0\!=\!0$. In the limit $ m_\phi t\ll 1$ we find $w(t)=C(z+m_\phi t)^{3/2}$. Putting $w(t_0)=\partial w(t_0)/\partial t=0$, see \eqref{y_comp} and \eqref{ampl_eq_syst}, one finally gets
\begin{equation}\label{y2_amp}
y_2(t)=(z+m_\phi t)^2.
\end{equation}
This solution ensures the amplitude dumping in a small vicinity of the
point \\$z+\sin m_\phi t_0=0$. Such behaviour matches with the oscillating picture: oscillations are suppressed when neutrino mass is vanishingly small.

%can be responsible for amplitude amplification during long coherent evolution of neutrino state.

\begin{figure}[!t]
	\centering
	\includegraphics[width=0.47\linewidth]{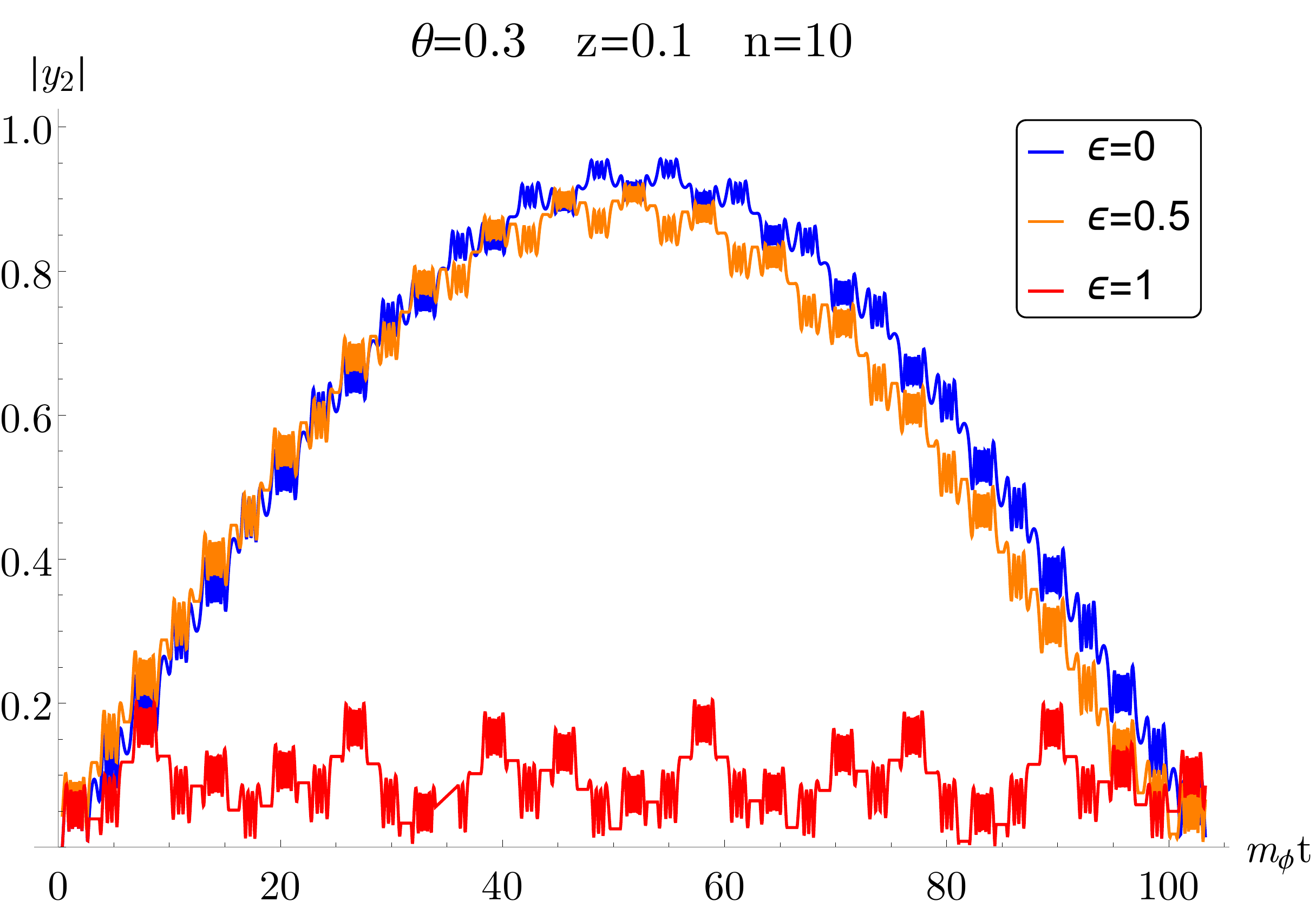}
	\includegraphics[width=0.47\linewidth]{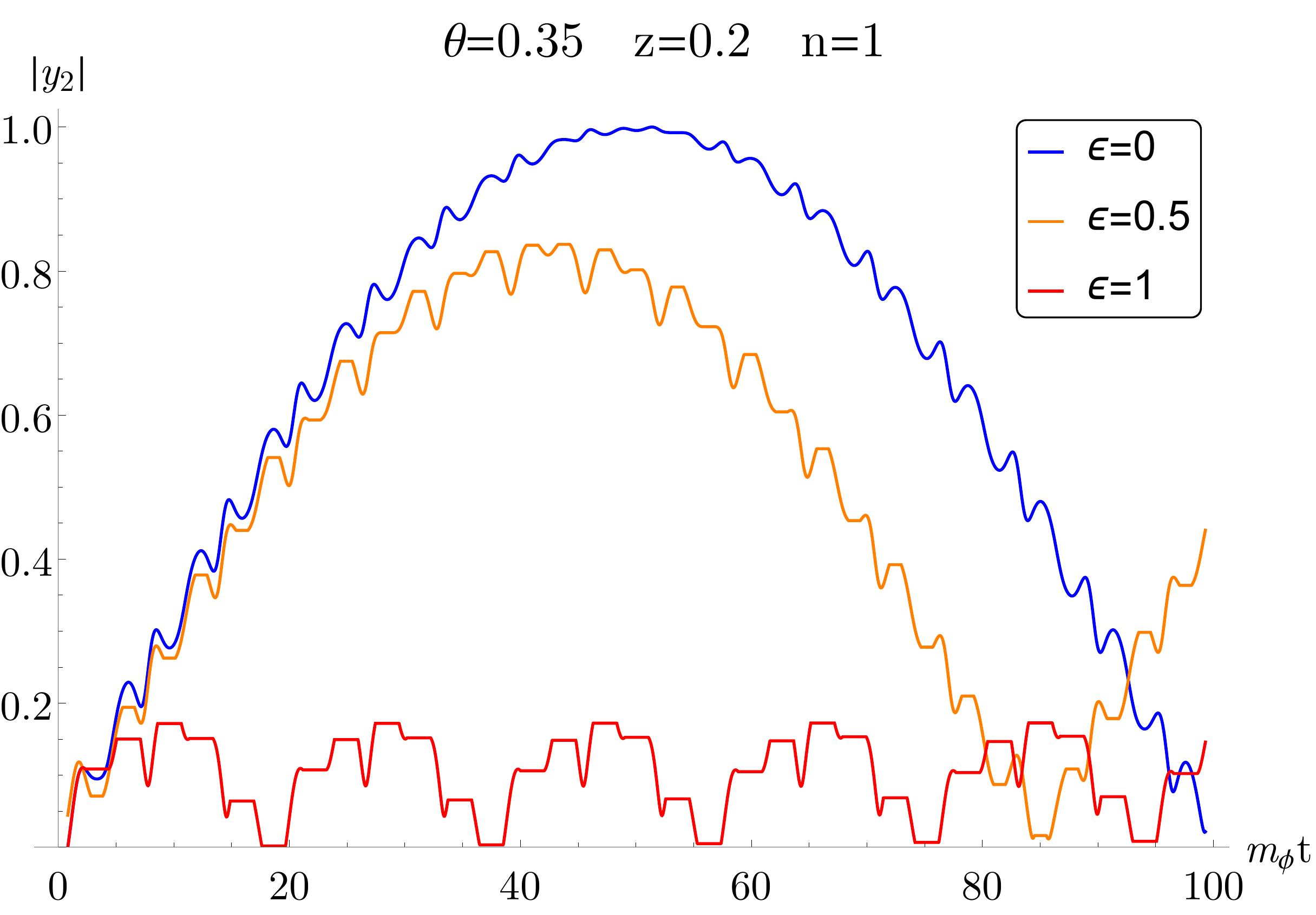}
	\caption{The oscillating behaviour of $|y_2(t)|$ in resonance \eqref{k_def} ($\epsilon=0$) and that excluding all regions near the points $M_\eff(t)=0$ \eqref{osc_not_dump} ($\epsilon=0.5,1$) for two sets of parameters.}
	\label{fig:2}
\end{figure}

Assuming \eqref{y2_osc} and \eqref{y2_amp} we make a reasonable assumption that \eqref{ampl_eq_SL} manifests non-trivial resonance behaviour near the transition between two regimes, \eqref{osc_vs_growth1} and \eqref{osc_vs_growth2}, at
%the amplitude changing regime manifests itself significantly only over  
\begin{equation}\label{osc_not_dump}
|z+\sin m_\phi t|\lesssim\epsilon\l\frac{m_\phi}{\beta}\r^{1/3}\quad{\rm where}\qquad \epsilon\sim1
\end{equation} 
%We recall that assumption $\beta\gg m_\phi$ makes condition~\eqref{osc_not_dump} sensible.
To verify this statement we resort to a numerical analysis. First, we solve
numerically \eqref{ampl_eq_syst} in resonance \eqref{k_def} and depict the resulting transition amplitude $|y_2(t)|$ for different model parameters in Fig.~\ref{fig:2} ($\epsilon=0$). To unmask a resonance dynamic in such a system we also solve \eqref{ampl_eq_syst} over long evolution period excluding the regions \eqref{osc_not_dump} where $y_1(t)$ and $y_2(t)$ were keeping constant. Results of such procedure for different values of $\epsilon$ are depicted on the same plots of Fig.~\ref{fig:2} ($\epsilon=0.5,1$).
%for two different sets of parameters which satisfy~\eqref{3:paramf_constr}
%The plots in Fig.~\ref{fig:2} are absolutely identical to those obtained in precise framework solving  equations on 2x2 density  matrix, see for details~\cite{Bezrukov:2018wvd}. It happens because~\eqref{paramf_constr} is assumed which justifies the approximate oscillation framework~\eqref{ampl_eq_syst} introduced in Sec.~\ref{Sec:first_order}.

We confirm that the regions $|z+\sin m_\phi t|\ll\l\frac{\disp m_\phi}{\disp
  \beta}\r^{1/3}$ \eqref{osc_vs_growth2} really give only sub-dominant
contributions to the transition probability as argued above. We also varified that our method in case $\epsilon\leq0.2$ enterily reproduces the resonant behaviour ($\epsilon=0$) but we do not show it in Fig.~\ref{fig:2} for a clear representation. For $\epsilon=1$ we do not find
any significant amplification of the amplitude $|y_2(t)|$. It
justifies our statement that non-trivial dynamic in system
\eqref{ampl_eq_syst} manifests itself mainly at \eqref{osc_not_dump}.

Once the relevant mass scale related to the resonance behaviour
\eqref{osc_not_dump} is found one can verify our primordial
assumptions \eqref{3:theta_constr}, \eqref{3:adiab_constr}. We assume
that the error budget of our framework \eqref{ampl_eq_syst} in the resonance
is insignificantly small if \eqref{3:theta_constr} and
\eqref{3:adiab_constr} are violated far beyond the region where
resonance reveals itself \eqref{osc_not_dump}. Consequently, we require
\begin{equation}\label{3:param_constr}
2\theta z\ll\l\frac{m_\phi}{\beta}\r^{1/3},\qquad\text{and}\qquad \l\theta z\frac{m_\phi}{2\beta}\r^{1/4}\ll\l\frac{m_\phi}{\beta}\r^{1/3}.
\end{equation}

Finally, our framework~\eqref{ampl_eq_syst} can be applied to describe
neutrino oscillations in resonance \eqref{k_def} only if \eqref{3:param_constr}, i.e., 
\begin{equation}\label{paramf_constr}
\begin{split}
2\theta z\ll&\l\frac{m_\phi}{\beta}\r^{1/3}\approx\frac{1}{n^{1/3}}
%m_\phi&\leqslant\beta
\end{split}
\end{equation}
is addressed where in last equality we used \eqref{3:z_small}. 

For the reasonable values of oscillations parameters \eqref{3:z_small}
and not too large $n$ the condition \eqref{paramf_constr} is always
fulfilled. This justifies the applicability of simplified oscillation
framework \eqref{ampl_eq_syst} used in the numerical analyses of Sec. \ref{Sec:resonance}.

 \bibliographystyle{JCAP-hyper}
\bibliography{note}

\providecommand{\href}[2]{#2}\begingroup\raggedright\begin{thebibliography}{10}

\bibitem{Bezrukov:2017ike}
F.~Bezrukov, A.~Chudaykin and D.~Gorbunov, \emph{{Hiding an elephant: heavy
  sterile neutrino with large mixing angle does not contradict cosmology}},
  \href{http://dx.doi.org/10.1088/1475-7516/2017/06/051}{\emph{JCAP} {\bf 1706}
  (2017) 051}, [\href{https://arxiv.org/abs/1705.02184}{{\tt 1705.02184}}].

\bibitem{Bezrukov:2018wvd}
F.~Bezrukov, A.~Chudaykin and D.~Gorbunov, \emph{{Induced resonance makes light
  sterile neutrino Dark Matter cool}},
  \href{https://arxiv.org/abs/1809.09123}{{\tt 1809.09123}}.

\bibitem{Bertone:2016nfn}
G.~Bertone and D.~Hooper, \emph{{History of dark matter}},
  \href{http://dx.doi.org/10.1103/RevModPhys.90.045002}{\emph{Rev. Mod. Phys.}
  {\bf 90} (2018) 045002}, [\href{https://arxiv.org/abs/1605.04909}{{\tt
  1605.04909}}].

\bibitem{Gorbunov:2014efa}
D.~S. Gorbunov, \emph{{Sterile neutrinos and their role in particle physics and
  cosmology}},
  \href{http://dx.doi.org/10.3367/UFNe.0184.201405i.0545}{\emph{Phys. Usp.}
  {\bf 57} (2014) 503--511}.

\bibitem{Bilenky:2016pep}
S.~Bilenky, \emph{{Neutrino oscillations: From a historical perspective to the
  present status}},
  \href{http://dx.doi.org/10.1016/j.nuclphysb.2016.01.025}{\emph{Nucl. Phys.}
  {\bf B908} (2016) 2--13}, [\href{https://arxiv.org/abs/1602.00170}{{\tt
  1602.00170}}].

\bibitem{Adhikari:2016bei}
M.~Drewes et~al., \emph{{A White Paper on keV Sterile Neutrino Dark Matter}},
  \href{http://dx.doi.org/10.1088/1475-7516/2017/01/025}{\emph{JCAP} {\bf 1701}
  (2017) 025}, [\href{https://arxiv.org/abs/1602.04816}{{\tt 1602.04816}}].

\bibitem{Abazajian:2017tcc}
K.~N. Abazajian, \emph{{Sterile neutrinos in cosmology}},
  \href{http://dx.doi.org/10.1016/j.physrep.2017.10.003}{\emph{Phys. Rept.}
  {\bf 711-712} (2017) 1--28}, [\href{https://arxiv.org/abs/1705.01837}{{\tt
  1705.01837}}].

\bibitem{Schneider:2016uqi}
A.~Schneider, \emph{{Astrophysical constraints on resonantly produced sterile
  neutrino dark matter}},
  \href{http://dx.doi.org/10.1088/1475-7516/2016/04/059}{\emph{JCAP} {\bf 1604}
  (2016) 059}, [\href{https://arxiv.org/abs/1601.07553}{{\tt 1601.07553}}].

\bibitem{Boyarsky:2008ju}
A.~Boyarsky, O.~Ruchayskiy and D.~Iakubovskyi, \emph{{A Lower bound on the mass
  of Dark Matter particles}},
  \href{http://dx.doi.org/10.1088/1475-7516/2009/03/005}{\emph{JCAP} {\bf 0903}
  (2009) 005}, [\href{https://arxiv.org/abs/0808.3902}{{\tt 0808.3902}}].

\bibitem{Gorbunov:2008ka}
D.~Gorbunov, A.~Khmelnitsky and V.~Rubakov, \emph{{Constraining sterile
  neutrino dark matter by phase-space density observations}},
  \href{http://dx.doi.org/10.1088/1475-7516/2008/10/041}{\emph{JCAP} {\bf 0810}
  (2008) 041}, [\href{https://arxiv.org/abs/0808.3910}{{\tt 0808.3910}}].

\bibitem{Roach:2019ctw}
B.~M. Roach, K.~C.~Y. Ng, K.~Perez, J.~F. Beacom, S.~Horiuchi, R.~Krivonos
  et~al., \emph{{NuSTAR Tests of Sterile-Neutrino Dark Matter: New Galactic
  Bulge Observations and Combined Impact}},
  \href{https://arxiv.org/abs/1908.09037}{{\tt 1908.09037}}.

\bibitem{Dodelson:1993je}
S.~Dodelson and L.~M. Widrow, \emph{{Sterile-neutrinos as dark matter}},
  \href{http://dx.doi.org/10.1103/PhysRevLett.72.17}{\emph{Phys. Rev. Lett.}
  {\bf 72} (1994) 17--20}, [\href{https://arxiv.org/abs/hep-ph/9303287}{{\tt
  hep-ph/9303287}}].

\bibitem{Shi:1998km}
X.-D. Shi and G.~M. Fuller, \emph{{A New dark matter candidate: Nonthermal
  sterile neutrinos}},
  \href{http://dx.doi.org/10.1103/PhysRevLett.82.2832}{\emph{Phys. Rev. Lett.}
  {\bf 82} (1999) 2832--2835},
  [\href{https://arxiv.org/abs/astro-ph/9810076}{{\tt astro-ph/9810076}}].

\bibitem{Farzan:2019yvo}
Y.~Farzan, \emph{{Ultra-light scalar saving the 3 + 1 neutrino scheme from the
  cosmological bounds}},
  \href{http://dx.doi.org/10.1016/j.physletb.2019.134911}{\emph{Phys. Lett.}
  {\bf B797} (2019) 134911}, [\href{https://arxiv.org/abs/1907.04271}{{\tt
  1907.04271}}].

\bibitem{Cline:2019seo}
J.~M. Cline, \emph{{Viable secret neutrino interactions with ultralight dark
  matter}},  \href{https://arxiv.org/abs/1908.02278}{{\tt 1908.02278}}.

\bibitem{Pavlinsky:2008ecy}
M.~Pavlinsky et~al., \emph{{Spectrum-Roentgen-Gamma astrophysical mission}},
  \href{http://dx.doi.org/10.1117/12.789192}{\emph{Proc. SPIE Int. Soc. Opt.
  Eng.} {\bf 7011} (2008) 70110H}.

\bibitem{Merloni:2012uf}
{\scshape eROSITA} collaboration, A.~Merloni et~al., \emph{{eROSITA Science
  Book: Mapping the Structure of the Energetic Universe}},
  \href{https://arxiv.org/abs/1209.3114}{{\tt 1209.3114}}.

\bibitem{Pusch:1982ps}
G.~D. Pusch, \emph{{NEUTRON OSCILLATIONS IN A PERIODICALLY VARYING MAGNETIC
  FIELD}}, \href{http://dx.doi.org/10.1007/BF02902503}{\emph{Nuovo Cim.} {\bf
  A74} (1983) 149}.

\bibitem{Egorov:1999ah}
A.~M. Egorov, A.~E. Lobanov and A.~I. Studenikin, \emph{{Neutrino oscillations
  in electromagnetic fields}},
  \href{http://dx.doi.org/10.1016/S0370-2693(00)01006-6}{\emph{Phys. Lett.}
  {\bf B491} (2000) 137--142},
  [\href{https://arxiv.org/abs/hep-ph/9910476}{{\tt hep-ph/9910476}}].

\bibitem{Dvornikov:2000mt}
M.~Dvornikov and A.~Studenikin, \emph{{Parametric resonance of neutrino
  oscillations in electromagnetic wave}},  in \emph{{New worlds in
  astroparticle physics. Proceedings, 3rd International Workshop, Faro,
  Portugal, September 1-3, 2000}}, pp.~126--131, 2000.
\newblock \href{https://arxiv.org/abs/hep-ph/0102099}{{\tt hep-ph/0102099}}.
\newblock \href{http://dx.doi.org/10.1142/9789812811035_0012}{DOI}.

\bibitem{Dvornikov:2004ms}
M.~Dvornikov, \emph{{Neutrino spin-flavor oscillations in frequently varying
  external fields}},
  \href{http://dx.doi.org/10.1134/S1063778807020159}{\emph{Phys. Atom. Nucl.}
  {\bf 70} (2007) 342--348}, [\href{https://arxiv.org/abs/hep-ph/0410152}{{\tt
  hep-ph/0410152}}].

\bibitem{Erdelyi}
A.~Erdelyi, \emph{{Asymptotic Expansions}}.

\bibitem{Abazajian:2001nj}
K.~Abazajian, G.~M. Fuller and M.~Patel, \emph{{Sterile neutrino hot, warm, and
  cold dark matter}},
  \href{http://dx.doi.org/10.1103/PhysRevD.64.023501}{\emph{Phys. Rev.} {\bf
  D64} (2001) 023501}, [\href{https://arxiv.org/abs/astro-ph/0101524}{{\tt
  astro-ph/0101524}}].

\bibitem{Sigl:1992fn}
G.~Sigl and G.~Raffelt, \emph{{General kinetic description of relativistic
  mixed neutrinos}},
  \href{http://dx.doi.org/10.1016/0550-3213(93)90175-O}{\emph{Nucl. Phys.} {\bf
  B406} (1993) 423--451}.

\bibitem{Shaposhnikov:2006xi}
M.~Shaposhnikov and I.~Tkachev, \emph{{The nuMSM, inflation, and dark matter}},
  \href{http://dx.doi.org/10.1016/j.physletb.2006.06.063}{\emph{Phys. Lett.}
  {\bf B639} (2006) 414--417},
  [\href{https://arxiv.org/abs/hep-ph/0604236}{{\tt hep-ph/0604236}}].

\bibitem{Kusenko:2006rh}
A.~Kusenko, \emph{{Sterile neutrinos, dark matter, and the pulsar velocities in
  models with a Higgs singlet}},
  \href{http://dx.doi.org/10.1103/PhysRevLett.97.241301}{\emph{Phys. Rev.
  Lett.} {\bf 97} (2006) 241301},
  [\href{https://arxiv.org/abs/hep-ph/0609081}{{\tt hep-ph/0609081}}].

\bibitem{Petraki:2007gq}
K.~Petraki and A.~Kusenko, \emph{{Dark-matter sterile neutrinos in models with
  a gauge singlet in the Higgs sector}},
  \href{http://dx.doi.org/10.1103/PhysRevD.77.065014}{\emph{Phys. Rev.} {\bf
  D77} (2008) 065014}, [\href{https://arxiv.org/abs/0711.4646}{{\tt
  0711.4646}}].

\bibitem{Malyshev:2014xqa}
D.~Malyshev, A.~Neronov and D.~Eckert, \emph{{Constraints on $3.55$ keV line
  emission from stacked observations of dwarf spheroidal galaxies}},
  \href{http://dx.doi.org/10.1103/PhysRevD.90.103506}{\emph{Phys. Rev.} {\bf
  D90} (2014) 103506}, [\href{https://arxiv.org/abs/1408.3531}{{\tt
  1408.3531}}].

\bibitem{Horiuchi:2013noa}
S.~Horiuchi, P.~J. Humphrey, J.~Onorbe, K.~N. Abazajian, M.~Kaplinghat and
  S.~Garrison-Kimmel, \emph{{Sterile neutrino dark matter bounds from galaxies
  of the Local Group}},
  \href{http://dx.doi.org/10.1103/PhysRevD.89.025017}{\emph{Phys. Rev.} {\bf
  D89} (2014) 025017}, [\href{https://arxiv.org/abs/1311.0282}{{\tt
  1311.0282}}].

\bibitem{Laine:2008pg}
M.~Laine and M.~Shaposhnikov, \emph{{Sterile neutrino dark matter as a
  consequence of nuMSM-induced lepton asymmetry}},
  \href{http://dx.doi.org/10.1088/1475-7516/2008/06/031}{\emph{JCAP} {\bf 0806}
  (2008) 031}, [\href{https://arxiv.org/abs/0804.4543}{{\tt 0804.4543}}].

\bibitem{Bezrukov:2009th}
F.~Bezrukov, H.~Hettmansperger and M.~Lindner, \emph{{keV sterile neutrino Dark
  Matter in gauge extensions of the Standard Model}},
  \href{http://dx.doi.org/10.1103/PhysRevD.81.085032}{\emph{Phys. Rev.} {\bf
  D81} (2010) 085032}, [\href{https://arxiv.org/abs/0912.4415}{{\tt
  0912.4415}}].

\bibitem{Kusenko:2010ik}
A.~Kusenko, F.~Takahashi and T.~T. Yanagida, \emph{{Dark Matter from Split
  Seesaw}}, \href{http://dx.doi.org/10.1016/j.physletb.2010.08.031}{\emph{Phys.
  Lett.} {\bf B693} (2010) 144--148},
  [\href{https://arxiv.org/abs/1006.1731}{{\tt 1006.1731}}].

\bibitem{Romanko}
V.~K. Romanko, \emph{{The course of differential equations and the calculus of
  variations}}.

\end{thebibliography}\endgroup
% \bibliography{FeebleScalar}

\end{document}